\documentclass[prd,preprint,preprintnumbers,nofootinbib,superscriptaddress]{revtex4-1}

\usepackage{epsfig}
\usepackage{amsbsy}
\usepackage{amssymb}
\usepackage{varioref}
\usepackage{pifont}

\usepackage[normalem]{ulem}
\usepackage[usenames]{color}

\def\sla#1{\ifmmode%
\setbox0=\hbox{$#1$}%
\setbox1=\hbox to\wd0{\hss$/$\hss}\else%
\setbox0=\hbox{#1}%
\setbox1=\hbox to\wd0{\hss/\hss}\fi%
#1\hskip-\wd0\box1 }
\newcommand{\half}{{\textstyle \frac{1}{2}}}

\newcommand{\twothird}{{\textstyle \frac{2}{3}}}

\begin{document}

\preprint{~MSUHEP--111114}

\title{Constraints on Little Higgs with Fully-Radiative Electroweak Symmetry Breaking}

\date{\today}

\author{Roshan Foadi}
\email{foadiros@msu.edu}
\affiliation{Service de Physique Th\'eorique, Universit\'e Libre de Bruxelles, 
Campus de la Plaine CP225, Bd du Triomphe, 1050 Brussels, Belgium }
\author{Carl R.~Schmidt}
\email{schmidt@pa.msu.edu}
\author{Jiang-Hao Yu}
\email{yujiangh@msu.edu}
\affiliation{Department of Physics and Astronomy,
Michigan State University, East Lansing, MI 48824, USA}

\begin{abstract}
In a recent paper, we introduced a new Little Higgs model, which contains the gauge structure $SU(2)^3\times U(1)$,
embedded in an approximate global $SO(5)\times SO(5)$ symmetry.
After breaking to the standard model, $SU(2)_L \times U(1)_Y$, this produces
two heavy $Z^\prime$ bosons and two heavy $W^{\prime\pm}$ bosons, along with a single
Standard Model-like Higgs scalar.  The unique feature of the model was that it was possible to obtain electroweak
symmetry breaking and a light Higgs mass entirely from perturbative loop contributions to the Higgs effective potential.
 In this paper we consider the electroweak constraints on this model, including tree and loop contributions to
 the universal oblique and non-oblique parameters, tree-level corrections to the $ZWW$ vertex, and
 tree and loop level corrections to $Zb\bar{b}$.  The most significant corrections are positive tree-level corrections
 to $\hat{S}$ and negative fermion-loop corrections to $\hat{T}$, which require that the scale for the global symmetry
 breaking be $\gtrsim2$ TeV, depending on the top-quark mixing parameter and the extra gauge couplings.  In addition, the loop corrections to
  $Zb\bar{b}$ contain a divergence that must be absorbed into the coefficient of a new operator in the theory.  The finite
  part of this $Zb\bar{b}$ correction, however, is negligible.
\end{abstract}

%\keywords{Beyond Standard Model , Spontaneous Symmetry Breaking}

\maketitle

%%%%%%%%%%%%%%%%%%%%%%%%%%%%%%%%%%

\section{Introduction}\label{sec:Intro}
We are on the threshold for discovering the mechanism of electroweak symmetry breaking (EWSB) and the origin of mass for the particles in
the standard model (SM). Data from the Large Hadron Collider (LHC) and the Tevatron will likely produce clues in the next few years
to help us unravel this mystery. Many possible scenarios have been considered for EWSB and the stabilization of the weak scale, with distinct phenomenology at colliders. In supersymmetry, for example, there is a symmetry between bosons and fermions which prevents quadratic divergences from destabilizing the Higgs potential. An intriguing feature of supersymmetry is radiative EWSB: loop effects drive the squared mass of the Higgs field to negative values, generating a nonzero vacuum expectation value (VEV). This is by no means an exclusive feature of supersymmetry. In fact non-supersymmetric scenarios with radiatively-induced EWSB are possible in the context of Little Higgs (LH) models~\cite{Georgi:1975tz}--\cite{amsterdam}  in which a set of approximate global symmetries are broken collectively, leaving a Higgs doublet to break the SM electroweak $SU(2)_L\times U(1)_Y$ gauge symmetry to the $U(1)_Q$ gauge symmetry of electromagnetism. It should be noted, however, that in ``standard'' LH models -- like the Minimal Moose~\cite{Arkani-Hamed:2002qx}, the Littlest Higgs~\cite{Arkani-Hamed:2002qy}, and the Simplest Little Higgs~\cite{simplest} -- EWSB depends crucially on the details of the ultraviolet (UV) completion of the theory. In fact certain effective operators, introduced to stabilize the potential, do not arise from radiative effects in these theories, but are rather added ``by hand'', as a remnant of UV dynamics.

In a recent paper~\cite{Foadi:2010bu}, we presented a Little Higgs model, based on the approximate $SO(5)\times SO(5)$ global symmetry, of which an $SU(2)^3\times U(1)$ subgroup is gauged.  In that paper we emphasized the unique feature of the model that the potential for the Higgs boson can be generated entirely through one-loop radiative corrections, and gives rise to a relatively light Higgs boson ($M_H\lesssim200$ GeV) even after effects at the UV cutoff scale are included. The model is built to minimally satisfy two requirements. First, as in all LH models, the Higgs doublet must be a pseudo-Goldstone boson of an approximate, spontaneously and collectively-broken global symmetry. Second, aside from Yukawa and hypercharge interactions, there should be no additional sources of custodial isospin violation. The fact that, as a byproduct, such a simple setup generates fully-radiative EWSB and a light Higgs, with little dependence on UV effects, is remarkable. The crucial feature of the model that allows this is an extended top quark sector that alone is sufficient to produce a
viable perturbative Higgs potential.

In the present paper and in a subsequent paper we shall focus on the phenomenology of this model, which is determined predominantly by the global and gauge symmetries.  Below the TeV scale, the only scalar boson is the Higgs boson, which acts pretty much like a SM Higgs boson.  This, in itself, is novel among Little Higgs models.  At the TeV scale there are two extra $SU(2)$ gauge triplets, {\it i.e.,} a pair of heavy $Z^\prime$ bosons and a pair of heavy $W^{\prime\pm}$ bosons.  This is the same as other $SU(2)^3\times U(1)$
extended gauge group models, such as a 4-site Higgsless model~\cite{Casalbuoni:2005rs,Accomando:2008dm} or a linear moose with multi-stage breaking of the $SU(2)$ gauge groups~\cite{Hsieh:2010zr}.  However, the unique way that the gauge symmetry is embedded into the approximate global symmetry gives new possibilities for the spectrum of heavy bosons, in addition to new electroweak constraints and collider phenomenology.  There are also new heavy fermion partners of the light quarks and leptons which may appear at the TeV scale.  In particular, the heavy partners of the top quark,
which are crucial to generating a viable Higgs potential, may have significant contributions to electroweak precision observables.

In the present paper we focus on the electroweak constraints on the model.
In Sec.~\ref{sec:Model} we review the model, paying particular attention to the gauge and fermion sector formulae that are needed to obtain the contributions to electroweak observables. In Sec.~\ref{sec:tree} we consider the tree level contributions that arise from the mixing of the
light and heavy gauge bosons. Here we extend the analysis of Ref.~\cite{Foadi:2010bu} on universal electroweak corrections by allowing a more general set of gauge parameters.  In addition, we consider the contributions to the $ZWW$ triple-gauge-boson vertex. In Sec.~\ref{sec:fermionloop} we analyze the contributions from the heavy partners of the top quark to
electroweak observables.  This includes loop corrections to the oblique parameters and the $Zb\bar{b}$ vertex,
as well as tree-level modifications to the top and bottom electroweak couplings.
Finally in Sec.~\ref{sec:conclusions} we present our conclusions.

\section{The Model}\label{sec:Model}
\begin{figure}[t]
\epsfig{file=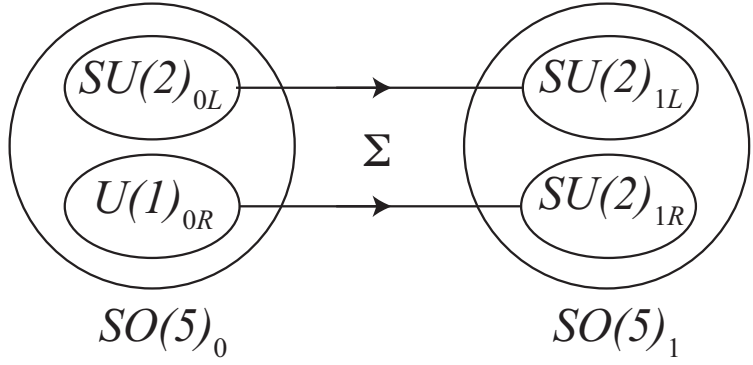,width=0.6\textwidth}
\caption{Moose diagram for the model.  The approximate global symmetry is
$SO(5)_0\times SO(5)_1$, with an embedded gauge symmetry of
$\left[SU(2)_{0L}\times U(1)_{0R}\right]\times\left[SU(2)_{1L}\times SU(2)_{1R}\right]$.
\label{fig:moose}}
\end{figure}
The symmetry structure\footnote{The model is motivated by the deconstruction of the 5-dimensional Gauge-Higgs
model of Ref.~\cite{Medina:2007hz}.  Models with related symmetry structure, but different fermion implementations,
are considered in the context of composite Higgs models in Refs.~\cite{Panico:2011pw,DeCurtis:2011yx}.}
of the model
 is represented in Fig.~\ref{fig:moose} by a moose diagram consisting of two sites, $0$ and $1$,
where the outer circles are the global $SO(5)$'s and the inner ellipses are the gauged subgroups.  In terms of the moose site indices,
the global symmetry can be written $SO(5)_0\times SO(5)_1$, while the gauged subgroup is $\left[SU(2)_{0L}\times U(1)_{0R}\right]\times\left[SU(2)_{1L}\times SU(2)_{1R}\right]$.  In this description the $L$ and $R$ subscripts
indicate the two commuting $SU(2)$ subgroups of $SO(5)$, while $U(1)_{0R}$ is a $U(1)$ subgroup of $SU(2)_{0R}$.
The global symmetry is spontaneously broken to the diagonal $SO(5)$ by a non-linear sigma field,
\begin{equation}
		\Sigma\ =\  e^{i\Pi/f}	\,,
\end{equation}
which transforms as $\Sigma\rightarrow U_0\Sigma U^\dagger_1$ under an $SO(5)_0\times SO(5)_1$
transformation.
Using the basis for the ten $SO(5)$ generator matrices, given in the Ref.~\cite{Foadi:2010bu}, we can  specify the goldstone boson fields as
\begin{eqnarray}
		\Pi \,=\, 					    \left(\begin{array}{ccccc}
					    \frac{\pi^3_L-\pi^3_R}{\sqrt{2}}&\pi^+_L&-\pi^-_R&0&\frac{v+h+i\pi^3}{\sqrt{2}}\\
					    \pi^-_L&\frac{-(\pi^3_L+\pi^3_R)}{\sqrt{2}}&0&-\pi_R^-&i\pi^-\\
					    -\pi^+_R&0&\frac{\pi^3_L+\pi^3_R}{\sqrt{2}}&\pi^+_L&i\pi^+\\
					    0&-\pi_R^+&\pi^-_L&\frac{-(\pi^3_L-\pi^3_R)}{\sqrt{2}}&\frac{v+h-i\pi^3}{\sqrt{2}}\\
					    \frac{v+h-i\pi^3}{\sqrt{2}}&-i\pi^+&-i\pi^-&\frac{v+h+i\pi^3}{\sqrt{2}}&0\\
					    		                             \end{array}\right)\,\ ,
\end{eqnarray}
The $\pi^a_L$ and $\pi^a_R$ are eaten by the heavy gauge fields and were set to zero (unitary gauge) in Ref.~\cite{Foadi:2010bu}.
Here we will need all of the goldstone boson interactions, so we keep them in the Lagrangian.

\subsection{Gauge Sector}\label{subsec:gauge}
\begin{figure}[t]
\epsfig{file=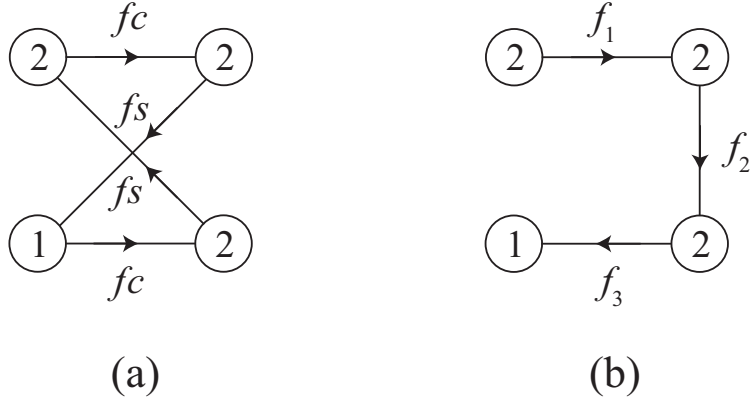,width=0.6\textwidth}
\caption{(a) $SU(2)^3\times U(1)$ moose, with equivalent gauge masses and mixings as our model.  (b)  Linear $SU(2)^3\times U(1)$ moose, such as the 4-site Higgsless model.
\label{fig:moosemass}}
\end{figure}

Gauging the  $\left[SU(2)_{0L}\times U(1)_{0R}\right]\times\left[SU(2)_{1L}\times SU(2)_{1R}\right]$ subgroup leads to the following covariant derivative
\begin{eqnarray}
		D^\mu \Sigma\  =\  \partial^\mu \Sigma
				- i g_{{0L}} W_{{0L}}^{a\mu} T_L^a \Sigma
				- i g_{{0R}} B_{{0R}}^{\mu} T_R^3 \Sigma
				+ i g_{{1L}} W_{{1L}}^{a\mu} \Sigma T_L^a
				+ i g_{{1R}} W_{{1R}}^{a\mu} \Sigma T_R^a	\,,
				\label{covderivs}
\end{eqnarray}
where, in the generator basis of Ref.~\cite{Foadi:2010bu},
\begin{eqnarray}
		T^a_L \  =\  \left(\begin{array}{cc}
						\Biggl( I\otimes\left(\half\sigma^a\right)\Biggr) &
						\begin{array}{c}
						0\\
						0\\
						0\\
						0\\
						\end{array} \\
						\begin{array}{cccc}
						\, 0\,  &\, 0\, &\, 0\, &\, 0\, \\
						\end{array}& 0 \\	
		                             \end{array}\right)\ , \qquad	
		T^a_R \  =\  \left(\begin{array}{cc}
						\Biggl(-\left(\half\sigma^a\right)^T\otimes I\Biggr) &
						\begin{array}{c}
						0\\
						0\\
						0\\
						0\\
						\end{array} \\
						\begin{array}{cccc}
						\ 0\  &\  0\  &\  0\  &\  0\ \\
						\end{array}& 0 \\	
		                             \end{array}\right)\ .      \nonumber
\end{eqnarray}
With this we can write the Lagrangian density for the gauge and sigma fields as
\begin{eqnarray}
{\cal L}_{\rm gauge}&=& -{1\over4}W_{0L}^{a\,\mu\nu}W^a_{0L\,\mu\nu}
-{1\over4}B_{0R}^{\mu\nu}B_{0R\,\mu\nu}-{1\over4}W_{1L}^{a\,\mu\nu}W^a_{1L\,\mu\nu}
-{1\over4}W_{1R}^{a\,\mu\nu}W^a_{1R\,\mu\nu}\nonumber\\
&&
+{f^2\over4}{\rm tr}\Big[\left(D^\mu\Sigma\right)\left(D_\mu\Sigma\right)^\dagger\Big]\ .
\label{eq:lagrange}
\end{eqnarray}
As shown in Ref.~\cite{Foadi:2010bu}, this leads to the gauge boson mass Lagrangian
\begin{eqnarray}
		{\cal L}_{\rm mass} &=& \frac{f^2}{4}\Biggl\{ g_{0L}^2 W_{0L}^{a\mu} W_{0L\mu}^{a}
				+  g_{0R}^2 B_{0R}^{\mu} B_{0R\mu}
				+  g_{1L}^2 W_{1L}^{a\mu} W_{1L\mu}^{a}
				+ g_{1R}^2 W_{1R}^{a\mu} W_{1R\mu}^{a}\nonumber\\
			&&\qquad-2c^2\, g_{0L} g_{1L} W_{0L}^{a\mu} W_{1L\mu}^{a}
			- 2s^2 \, g_{0L} g_{1R} W_{0L}^{a\mu} W_{1R\mu}^{a}\nonumber\\
			&&\qquad- 2s^2 \, g_{0R} g_{1L} B_{0R}^{\mu} W_{1L\mu}^{3}
			- 2c^2\, g_{0R} g_{1R} B_{0R}^{\mu} W_{1R\mu}^{3}\Biggr\}
	\,,\label{eq:gaugemass}
\end{eqnarray}
where\footnote{Please note that the definition of the variables $s$ and $c$ differs from that used in Ref.~\cite{Foadi:2010bu}
by a factor of $1/2$ in the argument of the sine and cosine.}
\begin{eqnarray}
		s\ =\ \sin\left(\frac{v}{2f}\right)\ ,\qquad\qquad c\ =\ \cos\left(\frac{v}{2f}\right)\ .
\end{eqnarray}
Throughout this paper, we shall use the parameter $s^2\approx v^2/(4f^2)$, as a small expansion parameter for comparison with the SM. It is interesting to note that this is the same mass Lagrangian that is obtained from the $SU(2)^3\times U(1)$ moose shown in Fig.~\ref{fig:moosemass}(a) (although without the extra triplet of uneaten goldstone bosons). Thus, the gauge boson spectrum and phenomenology will be distinct from other $SU(2)^3\times U(1)$ extended gauge group models, such as the 4-site Higgsless model~\cite{Casalbuoni:2005rs,Accomando:2008dm},
 that break via the linear moose shown in Fig.~\ref{fig:moosemass}(b).

The gauge boson masses and mixings were obtained in Ref.~\cite{Foadi:2010bu}.  Expressed as a power series in $s^2$,
we obtained two heavy gauge triplets with masses:
\begin{eqnarray}
		M_{W_L}^2&=&\half\left(g_{0L}^2+g_{1L}^2\right)f^2\ +\ \cdots\nonumber\\
		M_{Z_L}^2&=&\half\left(g_{0L}^2+g_{1L}^2\right)f^2\ +\ \cdots\nonumber\\
		M_{W_R}^2&=&\half g_{1R}^2f^2\ +\ \cdots\\
		M_{Z_R}^2&=&\half \left(g_{0R}^2+g_{1R}^2\right)f^2\ +\ \cdots\,,\nonumber
\end{eqnarray}
where the corrections are ${\cal O}(s^2)$.  The $W_L$ and $Z_L$ gauge bosons are very degenerate,
 since the mass splitting between these two states only arises at ${\cal O}(s^4)$.  The splitting between $W_R$ and $Z_R$
 depends strongly on the ratio of $g_{0R}/g_{1R}$.
The light SM $Z$ and $W$ bosons obtain masses
\begin{eqnarray}
		M_{W}^2&=& g_{L}^2f^2s^2\ +\ \cdots\nonumber\\
		M_{Z}^2&=&\left(g_{L}^2+g_{R}^2\right)f^2s^2\ +\ \cdots\,,
\end{eqnarray}
where the corrections are  ${\cal O}(s^4)$, and we have defined the couplings $g_L$ and $g_R$ by
\begin{eqnarray}
		\frac{1}{g_L^2}&=&\frac{1}{g_{0L}^2}+\frac{1}{g_{1L}^2}\nonumber\\
		\frac{1}{g_R^2}&=&\frac{1}{g_{0R}^2}+\frac{1}{g_{1R}^2}\,.\label{eq:gLgR}
\end{eqnarray}
The couplings $g_L$ and $g_R$, up to ${\cal O}(s^2)$ corrections, play the roles of the SM $SU(2)_L$ and $U(1)_Y$ gauge couplings, respectively.
Of course, the photon is exactly massless, being associated with the unbroken $U(1)_{EM}$, with coupling constant $e$ given
by
\begin{eqnarray}
		\frac{1}{e^2}&=&\frac{1}{g_{L}^2}+\frac{1}{g_{R}^2}\,=\,\frac{1}{g_{0L}^2}+\frac{1}{g_{1L}^2}+\frac{1}{g_{0R}^2}+\frac{1}{g_{1R}^2}\,.
\end{eqnarray}

\subsection{Fermion Sector}\label{subsec:fermion}
The fermions in the Little Higgs model are combined into multiplets of the global $SO(5)_{0}$ symmetry at site 0 only and transform under the corresponding gauge symmetries, $SU(2)_{0L}\times U(1)_{0R}$. For each generation of quarks in the SM, we have three multiplets of $SO(5)_0$, $\psi^A$, $\psi^B$, and $\psi^C$: the left-handed SM doublet is embedded in $\psi^A$, the right-handed up-type singlet is embedded in $\psi^B$, and the right-handed down-type singlet is embedded in $\psi^C$. The multiplets are Dirac multiplets, in that each comes in a right-handed and left-handed pair,
\begin{equation}
\psi \equiv \psi_L + \psi_R \ ,
\end{equation}
{\em except} that the SM fields within the multiplet are missing their Dirac partners\footnote{When applied to fermion fields, the subscripts $L$ and $R$ label the chirality. When applied everywhere else, they label the subgroup of $SO(5)$.}.

The $\psi^A$ field transforms as a 5 of $SO(5)$.  In terms of component fields, it  consists of
\begin{equation}
\psi_L^A\,=\, \left(\begin{array}{c}
				Q\\
				\chi\\
				u
		           \end{array}\right)^A_L\,, \quad
\psi_{R}^A\,=\, \left(\begin{array}{c}
				0 \\
				\chi\\
				u
		           \end{array}\right)^A_{R}\,,\label{eq:psia}
\end{equation}
where
\begin{equation}
Q\,=\, \left(\begin{array}{c}
				Q^{u} \\
				Q^d
		           \end{array}\right) \quad \mathrm{and} \quad
\chi\,=\, \left(\begin{array}{c}
				\chi^{y} \\
				\chi^u\\
		           \end{array}\right)\,
\end{equation}
transform as doublets under $SU(2)_{0L}$ and $u$ transforms as a singlet.  Note that the $Q_L$ field, which plays the role
of the light SM doublet field, is missing its partner $Q_R$ field.  Throughout this paper, we will use the
symbols $y$, $u$, and $d$ to indicate the electromagnetic charges of the fields by $q_{_{EM}}(y)=+5/3$, $q_{_{EM}}(u)=+2/3$,
and $q_{_{EM}}(d)=-1/3$.

The $\psi^B$ field also transforms as a 5 of $SO(5)$.  In terms of component fields, it  consists of
\begin{equation}
\psi_L^B\,=\, \left(\begin{array}{c}
				Q\\
				\chi\\
				0
		           \end{array}\right)^B_L\,,\quad
\psi_{R}^B\,=\, \left(\begin{array}{c}
				Q \\
				\chi\\
				u
		           \end{array}\right)^B_{R}\,.\label{eq:psib}
\end{equation}
The $u_R$ plays the role of the light SM field and is missing its partner $u_L$ field.

Finally, the $\psi^C$ field transforms as the adjoint 10 of $SO(5)$.  In terms of component fields, it  consists of
\begin{eqnarray}
\psi_{L}^C&=& \frac{1}{\sqrt{2}}\left(\begin{array}{ccccc}
				-u^-& \phi^{y} & 0 & 0 & Q^{u} \\
				\phi^{d} & -u^+ & 0  & 0 & Q^{d} \\
				-y & 0 & u^+  & \phi^y & \chi^{y} \\
				0 & -y & \phi^d & u^- & \chi^{u} \\
				\chi^{u} & -\chi^y & -Q^d  & Q^u & 0
		           \end{array}\right)^C_{L},\nonumber\\
		           \psi_{R}^C&=& \frac{1}{\sqrt{2}}\left(\begin{array}{ccccc}
				-u^-& \phi^{y} & -d & 0 & Q^{u} \\
				\phi^{d} & -u^+ & 0  & -d & Q^{d} \\
				-y & 0 & u^+  & \phi^y & \chi^{y} \\
				0 & -y & \phi^d & u^- & \chi^{u} \\
				\chi^{u} & -\chi^y & -Q^d  & Q^u & 0
		           \end{array}\right)^C_{R},
\end{eqnarray}
where
\begin{equation}
u^\pm\,=\,\frac{1}{\sqrt{2}}\left(u\pm\phi^u\right)\ ,
\end{equation}
and the SM right-handed singlet $d_R$ is missing its Dirac partner $d_L$. Under $SU(2)_{0L}$, the fields $\phi$ transform as triplets, the fields $Q$ and $\chi$ transform as doublets, and the fields $y$, $u$, and $d$ transform as singlets.

The Lagrangian density for the quark fields with Dirac masses can be written
\begin{eqnarray}
{\cal L}_{\rm Dirac}&=& i\bar{\psi}^A\sla{D}\psi^A-\lambda_Af\bar{\psi}^A\psi^A+i\bar{\psi}^B\sla{D}\psi^B-\lambda_Bf\bar{\psi}^B\psi^B\nonumber\\
&&
+\,i\,{\rm tr}\left(\bar{\psi}^C\sla{D}\psi^C\right)-\lambda_Cf{\rm tr}\left(\bar{\psi}^C\psi^C\right)
\ ,
\label{eq:lagrangebulk}
\end{eqnarray}
where the covariant derivatives are
\begin{eqnarray}
&&D^\mu\psi^{(A,B)}=\left[\partial^\mu
				- i g_{{0L}} W_{{0L}}^{a\mu} T_L^a
				- i g_{{0R}} B_{{0R}}^{\mu} \left(T^3_R+q_X\right) \right]\psi^{(A,B)}
\nonumber\\
&&D^\mu\psi^C=\partial^\mu \psi^C
				- i g_{{0L}} W_{{0L}}^{a\mu} \left[T_L^a, \psi^C\right]
				- i g_{{0R}} B_{{0R}}^{\mu} \left(\left[T^3_R,\psi^C\right]+q_X \psi^C\right)
\,.
\label{eq:covderivferm}
\end{eqnarray}
Under $U(1)_{0R}$ the fields transform with a charge given by $Y=T^3_R({\rm rep})+q_X$, where $q_X=+2/3$, and $T^3_R({\rm rep})$ is in the fundamental representation for $\psi^{A,B}$ and in the adjoint representation for $\psi^C$.  An equivalent Lagrangian for leptons can be constructed out of the same multiplets of fields, except with $q_X=0$.

With this Lagrangian all $\psi^A$ fields have a Dirac mass $M_A=\lambda_Af$, all
$\psi^B$ fields have a Dirac mass $M_B=\lambda_Bf$, and all $\psi^C$ fields have a Dirac mass $M_C=\lambda_Cf$,
{\em except} for the fields $Q_L$, $u_R$ and $d_R$, which are massless.  These are given mass, by introducing
the Yukawa terms,
\begin{eqnarray}
{\cal L}_{\rm Yukawa}&=& -\left[\lambda_1f\left(\bar{\psi}_{L}^A\Sigma \right) EE^\dagger\left(\Sigma^\dagger \psi_{R}^B\right)
+\sqrt{2}\lambda_2f\left(\bar{\psi}_{L}^A\Sigma\right) \left(1-EE^\dagger\right)\left(\Sigma^\dagger \psi_{R}^C\Sigma\right)E
+{\rm h. c.} \right]\nonumber\label{eq:lagrangebrane}\\
&=& -\left[\lambda_1f\left(\bar{\psi}_{L}^A\Sigma \right) EE^\dagger\left(\Sigma^\dagger \psi_{R}^B\right)
+\sqrt{2}\lambda_2f\left(\bar{\psi}_{L}^A\psi_{R}^C\Sigma\right)E+{\rm h. c.} \right]\ ,
\end{eqnarray}
where we have used the $O(4)$-invariant vector,
\begin{eqnarray}
		E \  =\ 	    \left(\begin{array}{rrrrr}
						0 \\
						0 \\
						0 \\
						0 \\
						1 \\	
		                             \end{array}\right)\ .
\end{eqnarray}
As explained in Ref.~\cite{Foadi:2010bu}, these Yukawa terms, along with the Dirac masses of Eq.~(\ref{eq:lagrangebulk}), maintain
the collective symmetry breaking necessary for the Little Higgs mechanism.  For $\lambda_{(1,2)}\ll\lambda_{(A,B,C)}$, they give
masses to the light SM fields of
\begin{eqnarray}
M_u&\approx&\lambda_1v/\sqrt{2}\nonumber\\
M_d&\approx&\lambda_2v/\sqrt{2}\label{eq:fermionmass}\,,
\end{eqnarray}
while the heavy fermions get only small shifts from their masses of $M_A$, $M_B$, $M_C$.

The only quark for which the approximation $\lambda_{1}\ll\lambda_{(A,B,C)}$ does not hold is the top quark.  Keeping
$\lambda_{1}$ for the top quark sector of the same order as $\lambda_{(A,B,C)}$ we find that the charge +2/3 fermions of $\psi_C$ and one
linear combination of each of the charge +2/3 fermions of $\psi_A$ and $\psi_B$ have mass eigenvalues unaffected by the Yukawa term.
The remaining three linear combinations mix due to the Yukawa term and have masses
\begin{eqnarray}
M_t^2&=&2\lambda_t^2f^2s^2\ +\ \cdots\nonumber\\
M_{T_A}^2&=&\left(\lambda_A^2+\lambda_{1}^{2}\right)f^2\ +\ \cdots\label{eq:topmass}\\
M_{T_B}^2&=&\lambda_B^2f^2\ +\ \cdots\nonumber\,,
\end{eqnarray}
where the corrections to $M_t^2$ are ${\cal O}(s^4)$, the corrections to $M_{T_{A,B}}^2$ are ${\cal O}(s^2)$ and we have defined
\begin{equation}
\frac{1}{\lambda_t^2}\,=\,\frac{1}{\lambda_{1}^{2}}+\frac{1}{\lambda_A^2}\,.\label{eq:lambdatop}
\end{equation}
We see that even for $\lambda_{1}$ not small, the top quark mass is down by a factor of $v/f$ compared to the heavy quarks.
More detailed expressions\footnote{Again, we alert the reader that the definitions of the parameters
$s$ and $c$ in this paper differ from those used in Ref.~\cite{Foadi:2010bu}, so care must be used in applying the formulae in that paper.}
 for the masses and mixings of the third generation charge +2/3 fermions are given in the Ref.~\cite{Foadi:2010bu}

As discussed in Ref.~\cite{Foadi:2010bu}, at the lowest order in the effective action, there are also fermion operators that correspond to renormalization of the broken currents of the complete $SO(5)$ multiplets:
\begin{eqnarray}
{\Delta\cal L}_{\rm Dirac}&=& i\kappa_A\bar{\psi}_L^A\left(\Sigma\sla{D}\Sigma^\dagger\right)\psi_L^A+ i\kappa_B\bar{\psi}_R^B\left(\Sigma\sla{D}\Sigma^\dagger\right)\psi_R^B\nonumber\\
&&
+\,i\kappa_{C_1}\,{\rm tr}\left[\bar{\psi}_R^C\left(\Sigma\sla{D}\Sigma^\dagger\right)\psi_R^C\right]+\,i\kappa_{C_2}\,{\rm tr}\left[\bar{\psi}_R^C\gamma^\mu\psi_R^C\left(D_\mu\Sigma\right)\Sigma^\dagger\right]\ .
\label{eq:kappas}
\end{eqnarray}
If we assume that all of the fermions act as fundamental point
particles, charged only under the $SU(2)_{0L}\times U(1)_{0R}$ gauge symmetry, then the $\kappa_i$ coefficients would arise only perturbatively through loop diagrams, and we
can assume them to be small.  In addition, for the coupling of the light SM fermions to the SM $Z$ and $W^\pm$
bosons, these operators have an additional suppression of $s^2$.  Thus, we will ignore these operators for the first two generations
of fermions.  However, it is possible that these coefficients may be larger for the third generation of fermions, so we will consider their effects on the SM currents of the top and bottom quarks.

\section{Tree Level Contributions to Precision Electroweak Observables}\label{sec:tree}
In our model there are contributions to precision electroweak observables at tree level, which arise from two separate sources. The first source can be identified by letting $\kappa_A=\kappa_B=\kappa_{C_1}=\kappa_{C_2}=0$. Then the SM light fermions couple directly only to the $SU(2)_{0L}\times U(1)_{0R}$ gauge fields, and the corrections to low-energy observables occur only through electroweak gauge current correlators, and are thus ``universal'' in the sense of Barbieri {\em et al.}~\cite{Barbieri:2004qk}. In Ref.~\cite{Foadi:2010bu}, we analyzed these tree-level contributions to electroweak precision constraints, but only put bounds on the parameters under the assumption $g_{1L}= g_{1R}$.  In this section we extend the analysis of the parameter space to include $g_{1L}\ne g_{1R}$. The second source of correction to precision electroweak observables comes from the $\kappa$-terms. These give corrections to low-energy observables of non-``universal'' type.  We shall postpone the discussion of these non-``universal'' corrections until section \ref{subsec:tbw}, where we consider their effects on the third generation quarks.

\subsection{``Universal'' Corrections}\label{subsec:treeoblique}
If  $\kappa_A=\kappa_B=\kappa_{C_1}=\kappa_{C_2}=0$, the electroweak gauge current correlators can be easily computed from the quadratic Lagrangian by inverting the subset of the propagator matrix involving the site-0 fields only. This leads to the following expressions for the electroweak parameters~\cite{Barbieri:2004qk}, to leading order in $s^2$:
\begin{eqnarray}
\Delta\hat{S}_{\rm tree} &=& s^2\left(\frac{g_L^2}{g_{1L}^2}\ +\ \frac{g_L^2}{g_{1R}^2}\right)\label{eq:S} \\
\Delta\hat{T}_{\rm tree} &=& 0 \label{eq:T} \\
\Delta Y_{\rm tree} &=& s^2\left(\frac{2g_L^2g_R^2}{g_{1R}^4}\right) \label{eq:Y} \\
\Delta W_{\rm tree} &=& s^2\left(\frac{2g_L^4}{g_{1L}^4}\right)\ . \label{eq:W}
\end{eqnarray}
To the order in $s^2$ that we are working in these equations, the couplings $g_{L}\equiv g$ and $g_{R}\equiv g^\prime$ are the SM
weak and hypercharge couplings, respectively, so that the electroweak observables depend only on three model parameters: $\{f,\,  g_{1L},\,  g_{1R}\}$.
Notice that the corrections to the electroweak observables are not oblique, since nonzero values for $Y$ and $W$ signal the presence of direct corrections, corresponding to four-fermion operator exchanges at zero momentum~\cite{Barbieri:2004qk,Chivukula:2004af}. Notice also that the custodial symmetry of the model ensures that $\hat{T}=0$ at tree-level. The global fit of the experimental data implies that a heavy Higgs boson is only compatible with positive $\hat{T}$~\cite{Barbieri:2004qk}. Therefore, we shall assume that the fermion parameters are chosen to obtain a light Higgs mass, as discussed in
Ref.~\cite{Foadi:2010bu}.

\begin{figure}[t]
\centerline{%
		\parbox{.45\textwidth}{\centerline{\epsfig{file=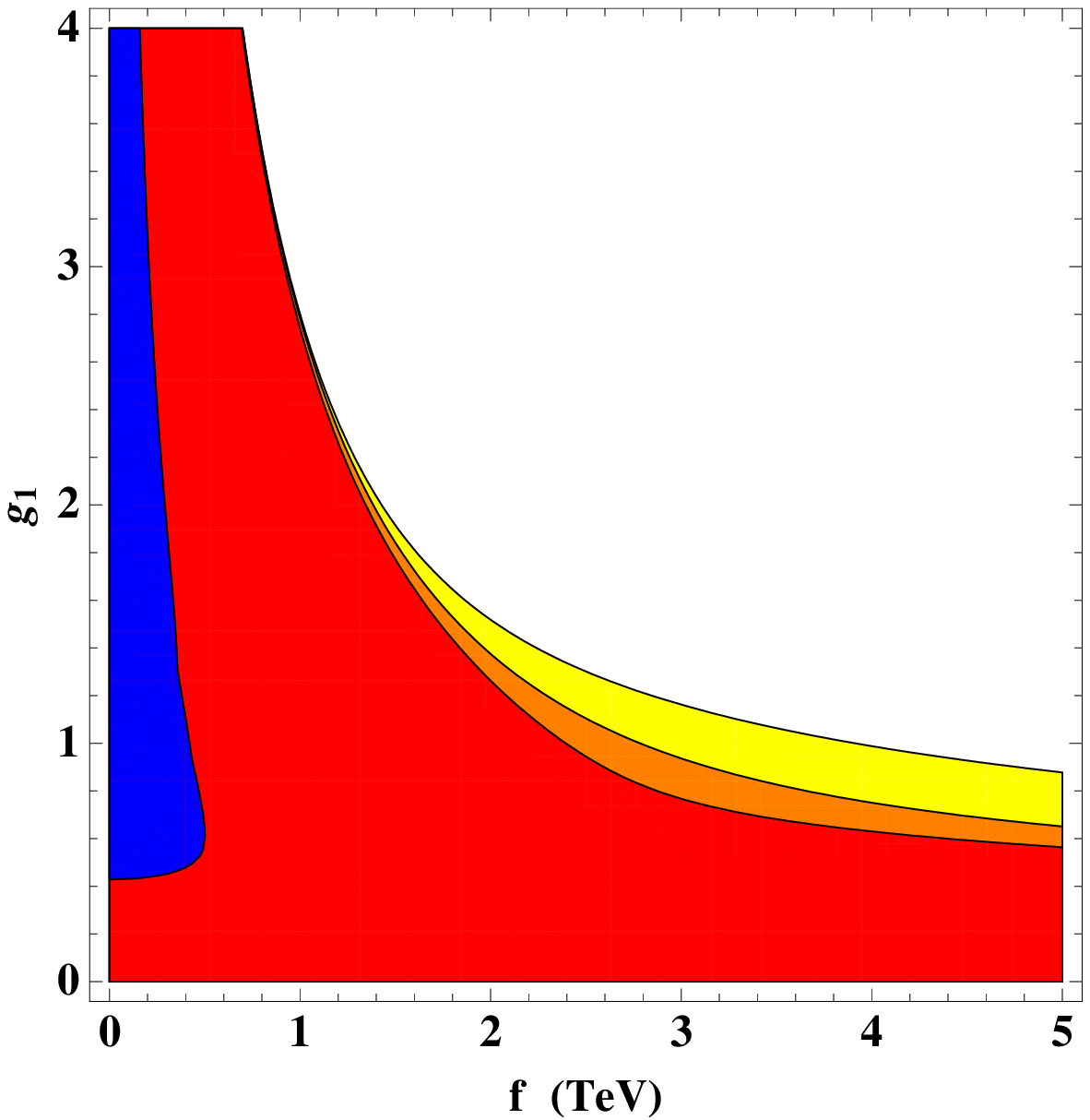,width=0.44\textwidth}}}~~~~
		\parbox{.45\textwidth}{\centerline{\epsfig{file=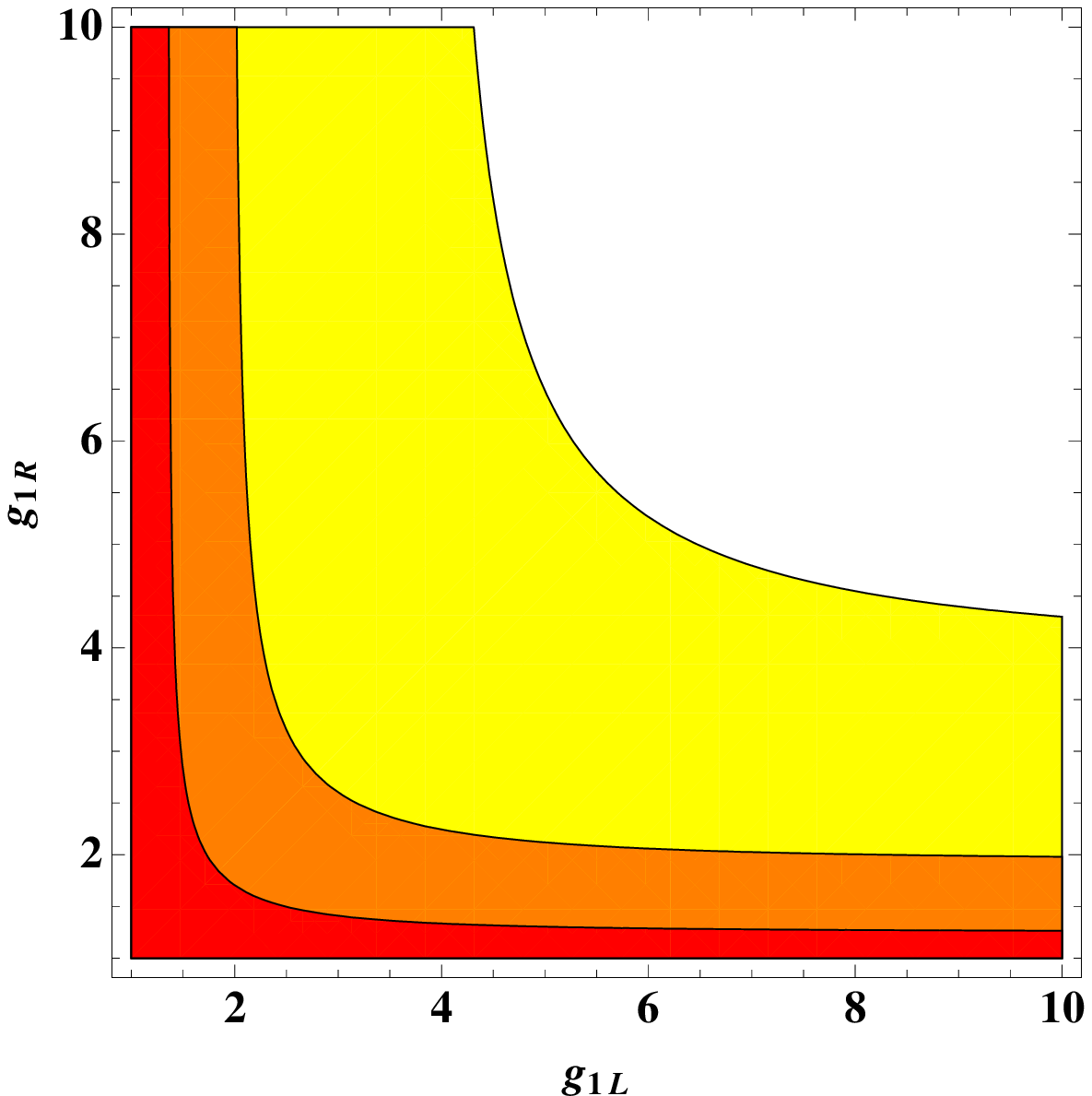,width=0.45\textwidth}}}}
\caption{
Constraints on $f$, $g_{1L}$, and $g_{1R}$ from $\hat{S}$, $\hat{T}$, $W$, and $Y$. The colored regions are excluded. Left: 95\% excluded regions in the $f$-$g_1$ plane, for different values of $\tan\phi$, where $g_1$ and $\tan\phi$ are defined in Eq.~(\ref{eq:g1phi}). The three main regions, from top to bottom, correspond to $\tan\phi=\infty$,  $\tan\phi=1$ (equivalent to $g_{1L}=g_{1R}=g_1$), and $\tan\phi=0$.  The blue region on the left side of the figure is the exclusion reach from experimental limits on the triple-gauge-boson parameter, $\Delta g_1^Z$,
at the 95\% confidence level.
Right: 95\% exclusion regions in the $g_{1L}$-$g_{1R}$ plane for $f=$0.5, 1.0, and 1.5 TeV, with the stronger constraints for the smaller values of $f$.\label{fig:ewtree}}
\end{figure}

The combined experimental constraints on $\hat{S}$, $\hat{T}$, $Y$, and $W$, taken from Ref.~\cite{Barbieri:2004qk}, for a light Higgs, give the bounds of Fig.~\ref{fig:ewtree}, where the colored regions are excluded. In Fig.~\ref{fig:ewtree} (left) we show the 95\% exclusion regions in the $f$-$g_1$ plane, for fixed values of $\tan\phi$, where we define
\begin{equation}
\frac{2}{g_1^2}\equiv\frac{1}{g_{1L}^2}+\frac{1}{g_{1R}^2}\ , \quad
\tan\phi\equiv \frac{g_{1R}}{g_{1L}} \ .
\label{eq:g1phi}
\end{equation}
The excluded regions, from top to bottom, correspond to $\tan\phi=\infty$, $\tan\phi=1$ (equivalent to $g_{1L}=g_{1R}=g_1$), and $\tan\phi=0$.
The weak dependence on $\tan\phi$ for small $f$ is due to the fact that the constraints are dominated by the contribution to $\hat{S}$,
which is independent of $\tan\phi$.  Only for smaller $g_1$ (and correspondingly larger $f$) are the
contributions to $Y$ and $W$ significant.
In Fig.~\ref{fig:ewtree} (right) the 95\% exclusion regions are shown in the $g_{1L}$-$g_{1R}$ plane for $f=$0.5, 1.0, and 1.5 TeV: of course, the smaller the value of $f$ the stronger the constraint.

\subsection{$ZWW$ vertex}
\label{subsec:ZWW}

Following Refs.~\cite{Hagiwara:1986vm,Chivukula:2006cg}, we can write the CP-invariant triple-gauge-boson vertex as
\begin{eqnarray}
{\cal L}_{\rm VVV}&=& ie\left[\left(W^{+\mu\nu}W^-_\mu-W^{-\mu\nu}W^+_\mu\right)A_\nu+\left(1+\Delta\kappa_\gamma\right)W^+_\mu W^-_\nu A^{\mu\nu}\right]\\
&&+ie\frac{c_Z}{s_Z}\left[\left(1+\Delta g_1^Z\right)\left(W^{+\mu\nu}W^-_\mu-W^{-\mu\nu}W^+_\mu\right)Z_\nu+\left(1+\Delta\kappa_Z\right)W^+_\mu W^-_\nu Z^{\mu\nu}\right]\ ,\nonumber
\label{eq:triplegaugeboson}
\end{eqnarray}
where
\begin{equation}
s_Z^2c_Z^2\ \equiv\ \frac{e^2}{4\sqrt{2}G_F M_Z^2}\ .
\end{equation}
In the SM at tree level,
$\Delta\kappa_Z=\Delta\kappa_\gamma=\Delta g_1^Z=0$. Using the Lagrangian of Eq.~(\ref{eq:lagrange}) and expanding in terms of the SM mass eigenstates, we obtain
\begin{eqnarray}
{\cal L}_{\rm WWV}&=& ie\left[\left(W^{+\mu\nu}W^-_\mu-W^{-\mu\nu}W^+_\mu\right)A_\nu+W^+_\mu W^-_\nu A^{\mu\nu}\right]\\
&&+ie\frac{\cos{\theta}}{\sin{\theta}}\left[\left(W^{+\mu\nu}W^-_\mu-W^{-\mu\nu}W^+_\mu\right)Z_\nu+W^+_\mu W^-_\nu Z^{\mu\nu}\right]\ +\
{\cal O}(s^4)\ ,\nonumber
\label{eq:triplegaugebosonii}
\end{eqnarray}
where $e=g_L\sin\theta=g_R\cos\theta$.
At first glance, this appears to give no correction to the SM triple-gauge-boson vertices at ${\cal O}(s^2)$; however, we must use the same
definition of the weak mixing angles in both equations.  We obtain
\begin{equation}
\frac{\cos{\theta}}{\sin{\theta}}\ =\ \frac{c_Z}{s_Z}\left[1-s^2\left(\frac{g_L^2+g_R^2}{g_L^2-g_R^2}\right)
\left(\frac{g_L^4}{g_{0L}^2g_{1L}^2}+\frac{g_R^4}{g_{0R}^2g_{1R}^2}\right)
\right]\ ,
\end{equation}
where $g_{0L}$ and $g_{0R}$ are determined in terms of $g_{1L}$ and $g_{1R}$ by the constraints of Eq.~(\ref{eq:gLgR}).
Comparing with the general form of the triple-gauge-boson vertices, Eq.~(\ref{eq:triplegaugeboson}), we obtain
\begin{equation}
\Delta\kappa_\gamma\ =\ 0\
\end{equation}
and
\begin{equation}
\Delta g_1^Z\ =\ \Delta\kappa_Z\ =\ -s^2\left(\frac{g_L^2+g_R^2}{g_L^2-g_R^2}\right)
\left[\frac{g_L^4}{g_{0L}^2g_{1L}^2}+\frac{g_R^4}{g_{0R}^2g_{1R}^2}\right]\ .
\end{equation}
Note that this expression is negative definite and the factor inside the brackets has a maximum value of $1/2$.

The LEP TGC Working group has obtained fits to the triple-gauge-boson vertex parameters~\cite{LEPTGC}, with the result at the 95\% confidence
level  of
\begin{equation}
-0.054\ <\ \Delta g_1^Z\ <\ +0.028\ ,
\end{equation}
under the assumption of $\Delta\kappa_\gamma=0$ and $\Delta\kappa_Z=\Delta g_1^Z$.  Using this result, we find that the constraints from the $ZWW$ vertex are always weaker than those from the universal electroweak parameters, as shown in the left plot of Fig.~\ref{fig:ewtree}.

\section{Contributions to Electroweak Corrections from the Top Quark Sector}
\label{sec:fermionloop}

\subsection{``Universal'' Corrections at One Loop}
\label{subsec:loopoblique}

\begin{figure}[t]
\epsfig{file=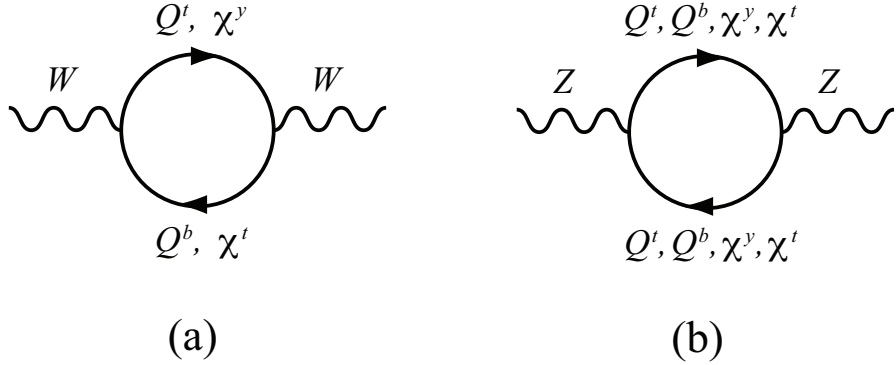,width=0.75\textwidth}
\caption{Feynman diagrams necessary to obtain corrections to $\hat{S}$ and $\hat{T}$ from the
third generation quarks.  Each pair of fields in the loop can come from either the $\psi^A$ or the $\psi^B$ $SO(5)$ multiplet.
\label{fig:SandTdiag}}
\end{figure}

The precision electroweak observables discussed in section \ref{sec:tree} also obtain corrections at the loop level in our model.
Just as in the SM, the dominant contributions are from oblique corrections due to loops of the third generation quarks, as well as their heavy partners. With respect to these, the one-loop contributions from the gauge sector are suppressed by $m_W^2/m_t^2\simeq 0.2$, and will therefore be neglected. Here we only include the contribution from the third generation quarks and their heavy partners.  We also assume that the loop corrections to $Y$ and $W$ beyond the SM are negligible, since they arise at higher order in the momentum expansion.
At ${\cal O}(s^0)$ the corrections are the same as the SM.  The deviations from the SM arise at
${\cal O}(s^2)$, and are obtained by evaluating the Feynman diagrams of Fig.~\ref{fig:SandTdiag}.  Note that two types
of $SU(2)_{0L}$ doublets are involved: $(Q^{t},Q^{b})$ and $(\chi^{y},\chi^{t})$, each coming from either the $\psi^A$ or $\psi^B$ $SO(5)$ multiplets.  (The non-singlet fields in $\psi^C$ always come in mass-degenerate doublets or triplets with vector couplings, and therefore, do not contribute to either $\hat{S}$ or $\hat{T}$.)
The charge +5/3 fields, $\chi^{y(A,B)}$, are mass eigenstates with masses $M_{A,B}$. The charge -1/3 field, $Q^{b(B)}$,
is a mass eigenstate with mass $M_B$.  The charge -1/3 field $Q^{b(A)}$ is the left-handed component of the SM bottom quark.  The charge +2/3 fields  $\left\{Q^{t(A,B)},\chi^{t(A,B)}\right\}$ mix, along with the singlets $t^{(A,B)}$, to give mass eigenstates, which are the SM top quark plus heavy partners.  The five mass eigenvalues are
\begin{eqnarray}
M_A^2 &=& \lambda_A^2 f^2 \nonumber\\
M_{T_A}^2 &\approx& (\lambda_A^2+\lambda_1^2) f^2 \nonumber\\
M_{T_B}^2 &\approx& M_{B}^2\ =\ \lambda_B^2 f^2 \\
M_{t}^2 &\approx& 2\lambda_t^2f^2s^2 \ ,\nonumber
\end{eqnarray}
where $M_t^2$ has a correction of ${\cal O}(s^4)$, $M_{T_A}^2$ and$M_{T_B}^2$ have corrections of ${\cal O}(s^2)$, and $M_A$ and $M_B$ are exact and independent of $s^2$.

Calculating the diagrams in Fig.~\ref{fig:SandTdiag}, we obtain
\begin{eqnarray}
\Delta\hat{T}_{\rm loop} &=& s^2\,N_C\frac{ \lambda_t^2}{(4\pi)^2}\,\frac{\lambda_t^2}{\lambda_A^2}
\left[1+\frac{\lambda_t^2}{\lambda_1^2}\right]
\left\{
-2\ln\frac{M_{T_A}^2}{M_t^2}+2\frac{\lambda_A^6}{\lambda_t^6}\ln\frac{M_{T_A}^2}{M_A^2}
-5\frac{\lambda_A^2}{\lambda_1^2}-2\frac{\lambda_A^4}{\lambda_1^4}
\right\}, \label{eq:Tloop}
\end{eqnarray}
and
\begin{eqnarray}
\Delta\hat{S}_{\rm loop} &=& s^2\,N_C\frac{g_L^2 }{(4\pi)^2}\left\{\frac{1}{9}\frac{\lambda_t^2}{\lambda_A^2}\left(
-4\ln\frac{M_A^2}{M_t^2}+1\right)
+\frac{1}{9}\frac{\lambda_t^4}{\lambda_A^4}\left(
2\ln\frac{M_{T_A}^2}{M_t^2}-5\right)
\right.\label{eq:Sloop} \\
&&\qquad\qquad\quad+\frac{1}{9}\left(2-\frac{\lambda_t^2}{\lambda_A^2}\right)F_1(M_{T_A},M_A)+\frac{\lambda_A^2}{\lambda_1^2}F_2(M_{T_A},M_A)\nonumber\\
&&\left.\qquad\qquad\quad+\frac{1}{9}\frac{\lambda_1^2}{\lambda_1^2+\lambda_A^2+\lambda_B^2}F_3(M_{T_A},M_B)
\right\}\ ,\nonumber
\end{eqnarray}
where $N_C=3$ is the number of colors, and we have defined the functions
\begin{eqnarray}
F_1(m_1,m_2) &=&\left[-\frac{3}{2x^3}+\frac{9}{2x}+7\right]\ln\frac{m_1^2}{m_2^2}+\frac{3}{x^2}-5\nonumber\\
F_2(m_1,m_2) &=&\left[\frac{1}{x^2}-1\right]\ln\frac{m_1^2}{m_2^2}-\frac{2}{x}\\
F_3(m_1,m_2) &=&\left[\frac{3}{2x^5}+\frac{6}{x^2}+\frac{9}{2x}\right]\ln\frac{m_1^2}{m_2^2}-\frac{3}{x^4}
-\frac{1}{x^2}-\frac{12}{x}\ ,\nonumber
\end{eqnarray}
with $x=(m_1^2-m_2^2)/(m_1^2+m_2^2)$.

\begin{figure}[t]
\centerline{%
		\parbox{.50\textwidth}{\centerline{\epsfig{file=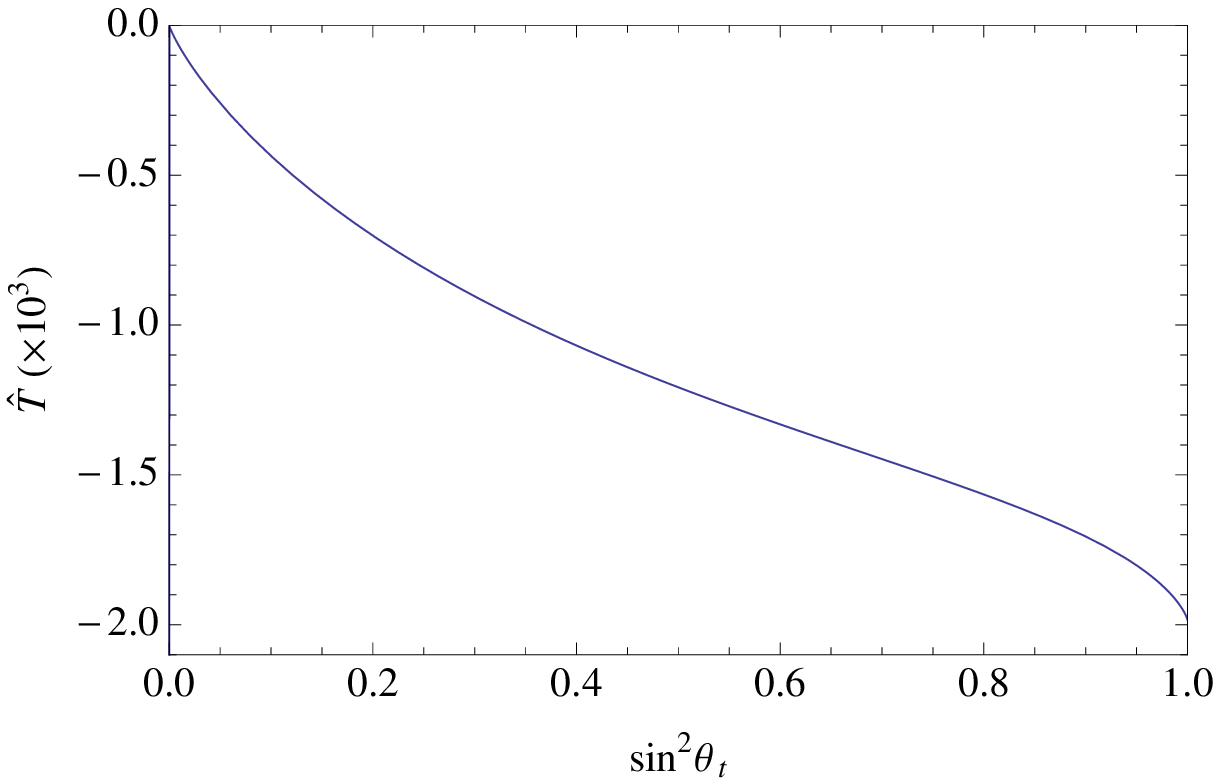,width=0.50\textwidth}}}~~~~
		\parbox{.50\textwidth}{\centerline{\epsfig{file=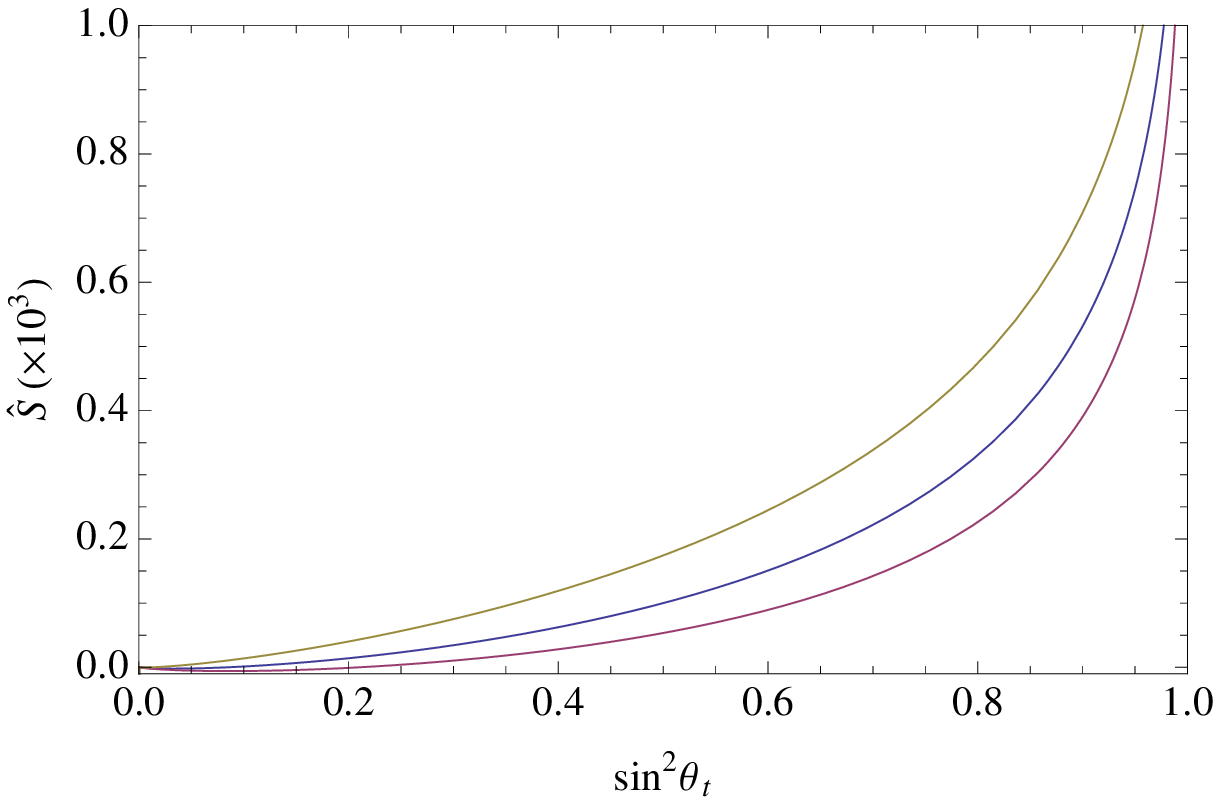,width=0.49\textwidth}}}}
\caption{
The heavy fermion contributions to $\hat{T}$ and $\hat{S}$ as a function of $\sin^2\theta_t$ for $f=1$ TeV.
The three curves of $\hat{S}$ from top to bottom correspond to
$\lambda_B=\lambda_A/2$, $\lambda_B=\lambda_A/\sqrt{2}$, and  $\lambda_B=\sqrt{2}\lambda_A$.
\label{fig:SandTplots}}
\end{figure}

In addition to dependence on the scale $f$, the loop corrections to $\hat{T}$  depend on the
parameters $\lambda_1$ and $\lambda_A$, while the loop corrections to $\hat{S}$ depend on
$\lambda_1$, $\lambda_A$, and $\lambda_B$.  However, the parameters $\lambda_1$ and $\lambda_A$
are not independent, being related to the SM top quark Yukawa, $\lambda_t$, by Eq.~(\ref{eq:lambdatop}).  Thus,
we can define a mixing angle in the top sector,
\begin{equation}
\sin\theta_t\ =\ \frac{\lambda_1}{\sqrt{\lambda_1^2+\lambda_A^2}}\ ,\label{eq:thetat}
\end{equation}
so that the top mass parameters are given in terms of $\theta_t$ by
$\lambda_A=\lambda_t/\sin\theta_t$ and $\lambda_1=\lambda_t/\cos\theta_t$.
Furthermore, $\lambda_B$ is determined in Ref.~\cite{Foadi:2010bu} in terms of the other parameters by the requirement
that the radiatively-generated Higgs potential has a minimum at $v=246$ GeV.   In Fig.~\ref{fig:SandTplots} we show
the loop contributions to $\hat{T}$ and $\hat{S}$ as a function of $\sin^2\theta_t$ for $f=1$ TeV.  For $\hat{S}$ we
choose the reasonable values of $\lambda_B=\lambda_A/2$, $\lambda_B=\lambda_A/\sqrt{2}$, and  $\lambda_B=\sqrt{2}\lambda_A$.

\begin{figure}[t]
\centerline{%
		\parbox{.30\textwidth}{\centerline{\epsfig{file=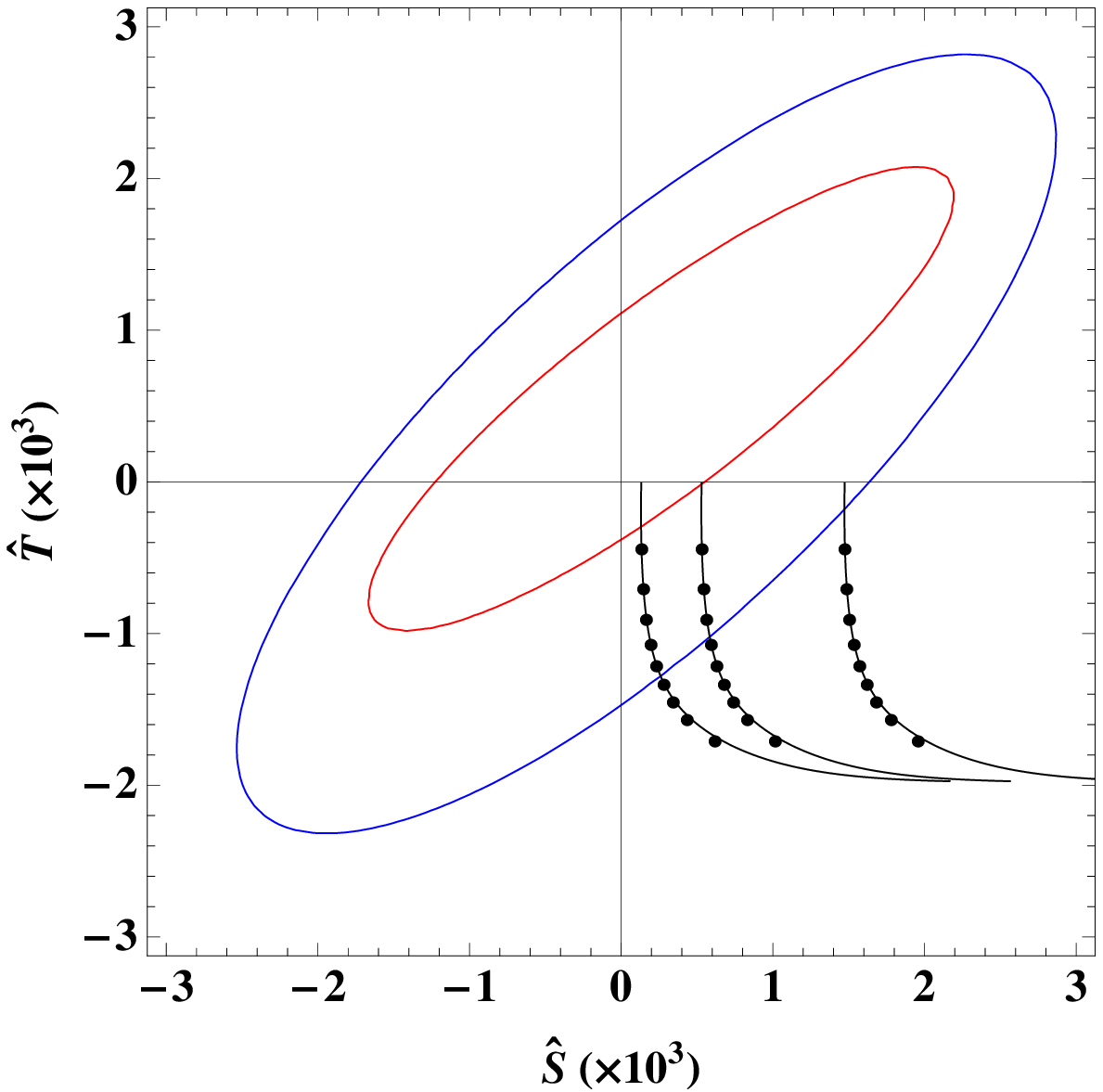,width=0.34\textwidth}}}~~~~
		\parbox{.30\textwidth}{\centerline{\epsfig{file=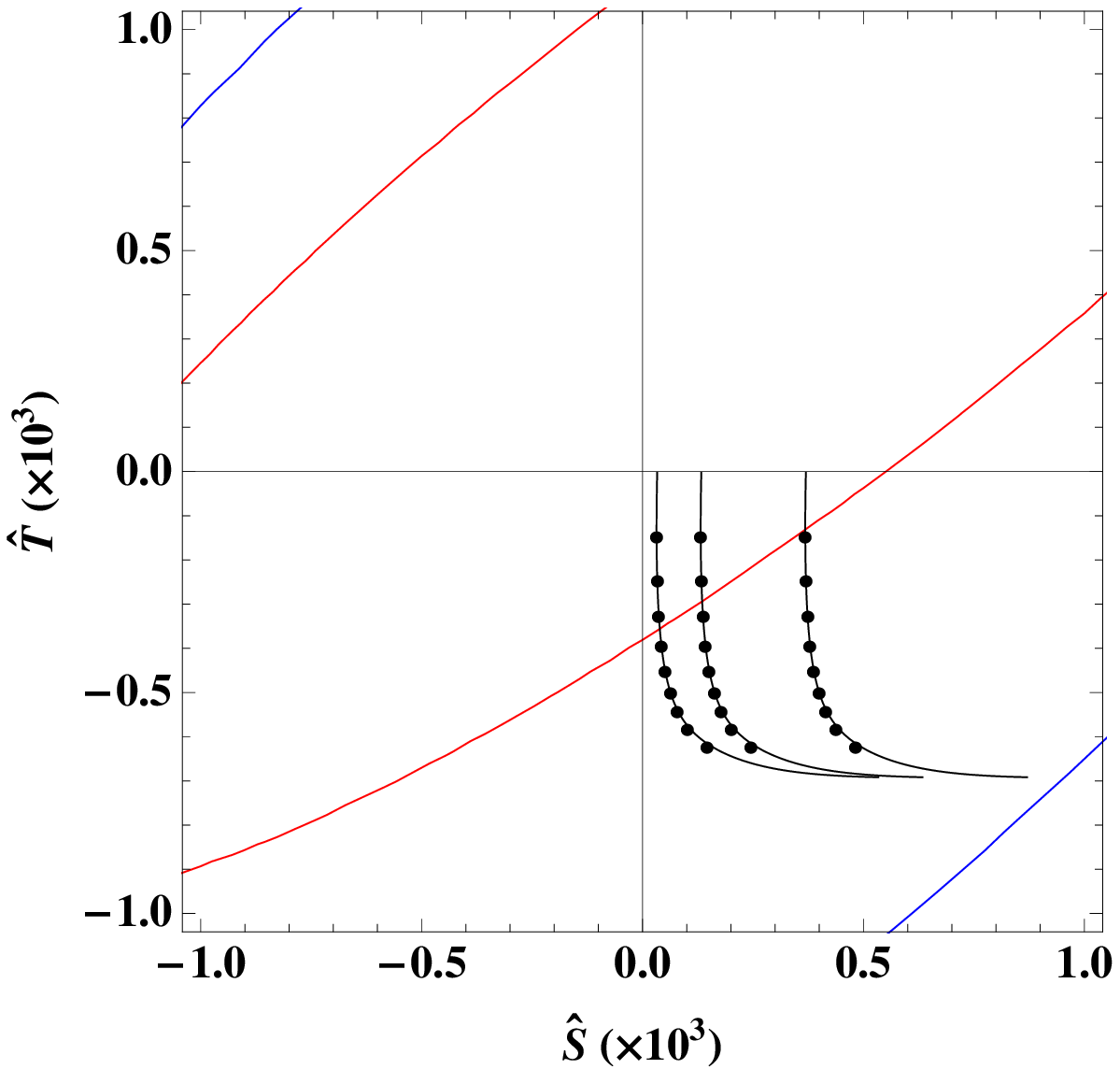,width=0.35\textwidth}}}
		\parbox{.35\textwidth}{\centerline{\epsfig{file=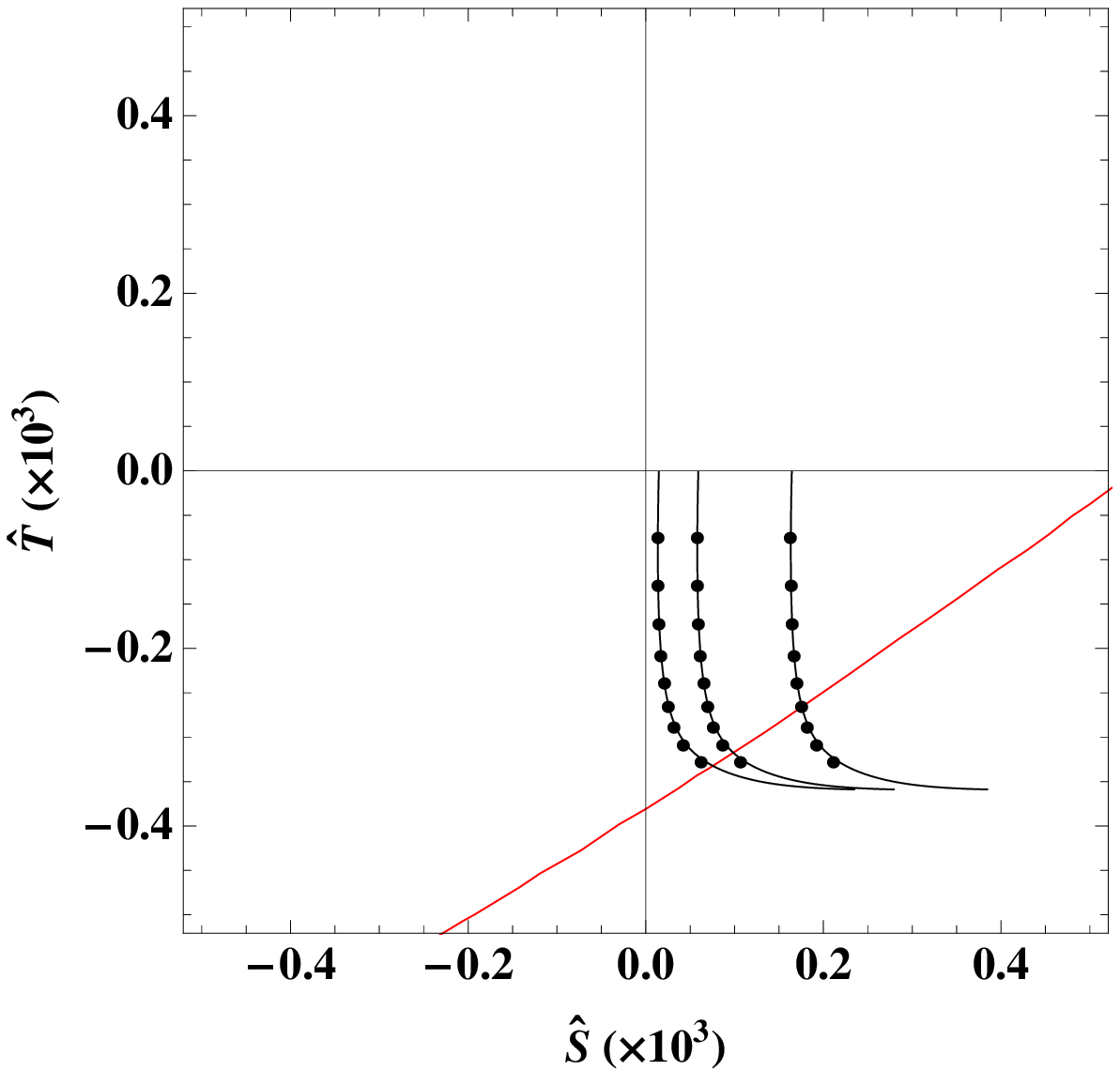,width=0.35\textwidth}}}}
\caption{Constraints on $f$, $g_1$ and $\sin^2{\theta_t}$ from $\hat{S}$ and $\hat{T}$, including both tree and loop  corrections. The left plot has $f=1$ TeV, the middle plot has $f=2$ TeV, and the right plot has $f=3$ TeV.
In each plot, the outer (blue) ellipse corresponds to the 95\% confidence levels (CL) contours from Ref.~\cite{Barbieri:2004qk}, assuming $Y=W=0$
and a light Higgs mass, while
the inner (red) ellipse corresponds to the 95\% CL contours from the Particle Data Group, Ref.~\cite{Nakamura:2010zzi},
using $\hat{S}=\alpha S/(4s_Z^2)$ and $\hat{T}=\alpha T$ with $U=0$ and $M_H=117$ GeV. The three curves in each plot are predictions for $\hat{S}$
and $\hat{T}$ for $0<{\sin^2\theta_t}<1$ with $g_1=10$, 5, and 3, as one moves to the right.  The dots on each curve
correspond to the values $\sin^2\theta_t=0.1, 0.2, 0.3, \dots, 0.9$, as one moves away from $\hat{T}=0$.  In the calculation of $\hat{S}$, we have fixed
$\lambda_B=\lambda_A/\sqrt{2}$.
\label{fig:STloop}}
\end{figure}

Of course, the loop corrections to $\hat{S}$ and $\hat{T}$ must be considered additively with the tree-level corrections from
section~\ref{sec:tree}.  Thus, the total corrections to $\hat{S}$ and $\hat{T}$ from the model will also depend on $g_{1L}$ and
$g_{1R}$, or alternatively, using Eq.~(\ref{eq:g1phi}), $g_1$ and $\tan{\phi}$.  We have seen in section \ref{sec:tree}
that the contribution of the non-oblique parameters $Y$ and $W$ become negligible for $g_1\gtrsim2$, and
consequently the dependence on $\tan\phi$ also becomes negligible.  Therefore, in this section we will fix $Y=W=0$
in our analysis and assume that the new physics only contributes to $\hat{S}$ and $\hat{T}$, which are then related
to the standard oblique parameters by $\hat{S}=\alpha S/(4s_Z^2)$ and $\hat{T}=\alpha T$ (and are independent of $\tan\phi$).  In addition, we will fix
$\lambda_B=\lambda_A/\sqrt{2}$, with the understanding that changes in this relationship will have
only a small effect on the results obtained below.  In Fig.~\ref{fig:STloop}
we plot the combined tree and loop-level calculations in the $\hat{S}$ and $\hat{T}$ plane as a function of $\sin^2\theta_t$
for various choices of $f$ and $g_1$, compared to limit contours from global fits.  The outer (blue)
ellipse in the plots corresponds to the 95\% confidence level (CL) contours from Ref.~\cite{Barbieri:2004qk}, assuming $Y=W=0$
and a light Higgs mass, while
the inner (red) ellipse corresponds to the 95\% CL contours from the Particle Data Group (PDG), Ref.~\cite{Nakamura:2010zzi} with $U=0$ and $M_H=117$ GeV.  The left plot has $f=1$ TeV, the middle plot has $f=2$ TeV, and the right plot has $f=3$ TeV.  The three curves in each plot are predictions for $\hat{S}$
and $\hat{T}$ for $0<{\sin^2\theta_t}<1$ with $g_1=10$, 5, and 3, as one moves to the right.  The dots on each curve
correspond to the values $\sin^2\theta_t=0.1, 0.2, 0.3, \dots, 0.9$, as one moves away from $\hat{T}=0$.

As seen in the left plot of Fig.~\ref{fig:STloop} for $f=1$ TeV, the loop contributions increase the tension with the
electroweak experimental data, due mostly to their negative contribution to $\hat{T}$.  Using the non-oblique universal fit of Ref.~\cite{Barbieri:2004qk}, requires $\sin^2\theta_t\lesssim0.04, 0.39, 0.55$ for $g_1=3, 5, 10$, respectively.  However, since for these values of $g_1$, we have found that $Y$ and $W$ are negligible, it seems more consistent to use the stronger, more up-to-date PDG limits from Ref.~\cite{Nakamura:2010zzi}.  In this case,
$g_1=3$  is ruled out at the 95\% CL, while $g_1=5$ and 10 are only allowed for very small $\sin^2\theta_t$.
For $f=2$ TeV, as seen in the middle plot, the PDG limits require $\sin^2\theta_t\lesssim0.09, 0.26, 0.32$ for $g_1=3, 5, 10$, respectively.  For $f=3$ TeV, as seen in the right plot, the PDG limits require $\sin^2\theta_t\lesssim0.60, 0.85, 0.91$ for $g_1=3, 5, 10$, respectively.

\begin{figure}[t]
	\epsfig{file=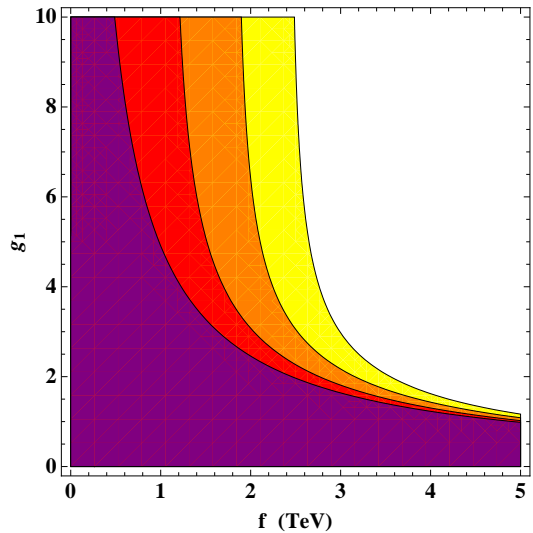,width=0.45\textwidth}
\caption{
Bounds on $f$ and $g_{1}$ from the PDG global fits to electroweak data~\cite{Nakamura:2010zzi}, compared to
tree-plus-loop universal corrections, where  $g_1$
is defined in Eq.~(\ref{eq:g1phi}).
The four regions are for $\sin^2\theta_t=0$, 0.1, 0.3,  and 0.6, with stronger constraints corresponding to larger $\sin^2\theta_t$.
\label{fig:looplimits}}
\end{figure}

We can use these results to find combined bounds on the parameters $f$, $g_1$ and $\sin^2\theta_t$.
Over most of the relevant range of $g_1$ we have found the tree-level contribution to $Y$ and $W$ to be
small; therefore, we shall neglect $Y$ and $W$ and use the PDG limits on $S$ and $T$ with $U=0$ and $M_H=117$ GeV.
In Fig.~\ref{fig:looplimits} the four excluded regions correspond to $\sin^2\theta_t=0$, 0.1, 0.3,  and 0.6, with stronger constraints for larger values of
$\sin^2\theta_t$.  From this plot we see that larger values of $\sin^2\theta_t$ require larger values of $f$ to avoid the electroweak constraints.  In Ref.~\cite{Foadi:2010bu} it was shown that increasing $f$ or decreasing $\sin^2\theta_t$
usually results in a larger mass for the Higgs boson (although cut-off effects in the fermion sector spoil a direct prediction
for $M_H$ in terms of these variables.)  Unfortunately, in the electroweak fits, larger values of $M_H$ move the
95\% CL ellipse in the $ST$-plane towards the upper left, which further increases the tension in this model.

The negative loop corrections to $\hat{T}$, as well as the smaller positive loop corrections to $\hat{S}$, appear to
be a generic prediction~\cite{SekharChivukula:2009if,Chivukula:2011jh} of models that feature a bi-doublet of $SU(2)_L\times SU(2)_R$.
In fact, our results for $\hat{S}$ and $\hat{T}$ are very similar to those found in Refs.~\cite{SekharChivukula:2009if,Chivukula:2011jh},  and agree  with Equations (41) and (43) from Ref.~\cite{Chivukula:2011jh} in certain limits of the parameters.
It is not unreasonable to expect similar significant loop corrections to $\hat{T}$ and $\hat{S}$ in other
models with the same symmetry structure, but different fermion implementations, such as Refs.~\cite{Panico:2011pw,DeCurtis:2011yx}.  Certainly, the tree level corrections from mixing with heavy gauge bosons would be the same as in our model.

\subsection{Tree Level Corrections to Top and Bottom Couplings to SM Gauge Bosons}
\label{subsec:tbw}

The top and bottom quark couplings to the $W^\pm$ and $Z$ bosons have corrections of  ${\cal O}(s^2)$, relative to that of the light
SM fermions, due to mixing with the heavy fermions in our model.  Since the couplings in the top quark sector are large,
we also consider the possibility that the
operators corresponding to renormalization of the broken currents can be non-negligible for the third generation.

For the correction to the top-bottom charged current, relative to that for the light quarks, we obtain
\begin{eqnarray}
\Delta{\cal L}_{\rm CC}&=& \frac{1}{\sqrt{2}}\bar{t}_L\gamma_\mu b_L
\Biggl\{-s^2\left[\frac{\lambda_t^4}{\lambda_A^4}g_{0L}W_{0L}^{+\mu}+\kappa_A\left(1-2\frac{\lambda_t^2}{\lambda_A^2}\right)g_{1L}W_{1L}^{+\mu}-\kappa_Ag_{1R}W_{1R}^{+\mu}\right]\nonumber\\
&&\qquad\qquad\quad\
+\kappa_A\left(g_{1L}W_{1L}^{+\mu}-g_{0L}W_{0L}^{+\mu}\right)\Biggr\}+{\rm h.c.\,,}\label{eq:CC}
\end{eqnarray}
where  we have only kept terms that couple to the $W^\pm$ and $Z$ bosons to at most ${\cal O}(s^2)$. (Note that the
term on the second line affects the couplings to the heavy $W_{L,R}^\pm$ bosons at ${\cal O}(s^0)$, but affects
the couplings to the $W^\pm$ only at ${\cal O}(s^2)$.    )  The terms that are proportional to factors of $\lambda_t/\lambda_A$ arise from mixing in the top quark sector.
Expanding the gauge bosons in terms of the mass eigenstates, we find the correction to the $tbW$ current at ${\cal O}(s^2)$ to be
\begin{eqnarray}
\Delta{\cal L}_{ tbW}&=&\frac{g_L}{\sqrt{2}}\bar{t}_L\gamma_\mu b_L W^{+\mu}\left\{-s^2
\left[\frac{\lambda_t^4}{\lambda_A^4}+\kappa_A\left(1-2\frac{\lambda_t^2}{\lambda_A^2}+\frac{g_L^2}{g_{0L}^2}-\frac{g_L^2}{g_{1L}^2}\right)\right]\right\}+{\rm h.c.\,.}\label{eq:tbW}
\end{eqnarray}
We see from this result that there are no right-handed charged couplings for the third-generation SM fields in this model.
This is easily understood from the fact that $b_R$ lies in the $\psi^C$ multiplet, but none of the charge $+2/3$ quarks in this
multiplet mix with the SM top quark (in the limit of zero bottom quark mass).
Based on absence of right-handed charged couplings and using the results of
Ref.~\cite{Larios:1999au}, we expect that constraints from $b\rightarrow s\gamma$ on the parameters in our model to be negligible.  The left-handed $tbW$ couplings can be probed in the single-top-quark production process~\cite{Berger:2009hi,AguilarSaavedra:2008gt}; however, since the current statistics from CDF and D0 are small, the resulting constraints
on the parameters in our model are weak.

For the correction to the top quark neutral current, relative to that for the up quark, we obtain to ${\cal O}(s^2)$,
\begin{eqnarray}
\Delta{\cal L}_{t\bar{t}\rm NC}&=& \half\bar{t}_L\gamma_\mu t_L
\Biggl\{-2s^2\left[\left(g_{0L}W_{0L}^{3\mu}-g_{0R}B_{0R}^{\mu}\right)\frac{\lambda_t^4}{\lambda_A^4}
+\kappa_A\left(1-2\frac{\lambda_t^2}{\lambda_A^2}\right)\left(g_{1L}W_{1L}^{3\mu}-g_{1R}W_{1R}^{3\mu}\right)\right] \nonumber\\
&& \qquad\qquad\quad+
\kappa_A\left(g_{1L}W_{1L}^{3\mu}-g_{0L}W_{0L}^{3\mu}+g_{0R}B_{0R}^{\mu}-g_{1R}W_{1R}^{3\mu}\right)\Biggr\} \nonumber\\
&&+\half\bar{t}_R\gamma_\mu t_R
\left\{-2s^2\,\frac{\lambda_t^2}{\lambda_A^2}\left(g_{0L}W_{0L}^{3\mu}-g_{0R}B_{0R}^{\mu}\right)\right\}.\label{eq:NCt}
\end{eqnarray}
Expanding the gauge bosons in terms of the mass eigenstates, we find the correction to the $Zt\bar{t}$ current at ${\cal O}(s^2)$ to be
\begin{eqnarray}
\Delta{\cal L}_{ Zt\bar{t}}&=& \frac{e}{\sin{2\theta}}\bar{t}_L\gamma_\mu t_L Z^\mu
\left\{-2s^2\left[\frac{\lambda_t^4}{\lambda_A^4}
+\kappa_A\left[\left(1-2\frac{\lambda_t^2}{\lambda_A^2}\right)+\frac{1}{2}\left(\frac{g_L^2}{g_{0L}^2}-\frac{g_L^2}{g_{1L}^2}+\frac{g_R^2}{g_{0R}^2}-\frac{g_R^2}{g_{1R}^2}\right)\right]\right] \right\}\nonumber\\
&&+\frac{e}{\sin{2\theta}}\bar{t}_R\gamma_\mu t_R Z^\mu
\left\{-2s^2\,\frac{\lambda_t^2}{\lambda_A^2}\right\}.\label{eq:ttZ}
\end{eqnarray}
The current bounds on the $Zt\bar{t}$ coupling from collider physics is very weak. Although the top pair production from an off-shell $Z$ exchange is relevant to the $Zt\bar{t}$ coupling measurement,  at the  Tevatron and LHC it is very hard to extract the EW contributions through the $Zt\bar{t}$ couplings from the huge rates from the QCD production of the top pair.

For the correction to the bottom quark neutral current, relative to that for the down quark, we obtain to ${\cal O}(s^2)$,
\begin{eqnarray}
\Delta{\cal L}_{b\bar{b}\rm NC}&=& \half\bar{b}_L\gamma_\mu b_L
\,\kappa_A\left(g_{0L}W_{0L}^{3\mu}-g_{1L}W_{1L}^{3\mu}+g_{0R}B_{0R}^{\mu}-g_{1R}W_{1R}^{3\mu}\right)\label{eq:kappabb}\\
&&+\half\bar{b}_R\gamma_\mu b_R \left(\kappa_{C_1}+\kappa_{C_2}\right)\left[g_{0R}B_{0R}^{\mu}-g_{1R}W_{1R}^{3\mu}-s^2\left(g_{1L}W_{1L}^{3\mu}-g_{1R}W_{1R}^{3\mu}\right)\right].\label{eq:NCb}\nonumber
\end{eqnarray}
In this case the only large corrections are due to current renormalizations proportional to $\kappa_i$, since the mixing of the bottom
quark with the heavy charged -1/3 fermions is suppressed by factors of $\lambda_2\propto \sqrt{2}m_b/v$.  Furthermore, we expect
the correction to the right-handed $Zb\bar{b}$ current, which is proportional to $\kappa_{C_1}+\kappa_{C_2}$, to be similarly suppressed,
because the mass mixing of the $\psi^C$ fields (which are the fermion fields that are involved in the $\kappa_{C_{1,2}}$ operators) are also suppressed by factors of $\lambda_2$.  On the other hand, the $\psi^A$ fields have mass mixings proportional to $\lambda_1$, which is ${\cal O}(1)$ for the top quark, so the corrections to the left-handed $Zb\bar{b}$ current proportional to $\kappa_A$ may be significant.
This argument will be borne out in the perturbative calculation in the next section.

Finally, we note that the coefficient $\kappa_B$ does
not come into play at all for the corrections to the currents involving the SM fermions.  This is due to the fact that $t_R$, which is the SM field lying in $\psi^B$, is a singlet under both $SU(2)_L$ and $SU(2)_R$.

We shall postpone the discussion of the phenomenological constraints on the $Zb\bar{b}$ vertex until the next section, where we perform
the corrections to this vertex at one loop.

\subsection{Loop Corrections to the $Zb\bar{b}$ Vertex}
\label{subsec:Zbb}

In the SM the one-loop electroweak corrections to the $Zb\bar{b}$ vertex are dominated by the contribution proportional to
$(m_t/v)^2$.  In our model there will be additional corrections, which are suppressed by powers of $s^2$ relative to the SM
result.  In this section we calculate these ${\cal O}(s^2)$ corrections in the limit of zero electroweak gauge couplings and zero bottom
quark mass.

Before discussing the loop calculation, it is useful to
re-diagonalize the Goldstone bosons from the fields ($\pi^a$, $\pi_L^a$, $\pi_R^a$) into three new sets of triplets
($\phi^a$, $\rho^a$, $\eta^a$) by:
\begin{eqnarray}
\eta^a&=&\frac{\pi_L^a+\pi_R^a}{\sqrt{2}}\nonumber\\
\phi^a&=&N\left[c\,\pi^a+s\left(\frac{\pi_L^a-\pi_R^a}{\sqrt{2}}\right)\right]\nonumber\\
\rho^a&=&N\left[-s\,\pi^a+c\left(\frac{\pi_L^a-\pi_R^a}{\sqrt{2}}\right)\right]
\ ,\nonumber
\end{eqnarray}
where the normalization factor,
\begin{equation}
N\,=\,\frac{s}{(v/2f)}\  ,\end{equation}
 ensures that the kinetic terms for the new triplets have the standard normalization after
expanding $\Sigma$ to all orders in $v/f$.
This choice of fields is convenient, because $\phi^\pm$ are the only charged scalars that couple to the bottom quark in a three-point vertex, a fact that greatly simplifies the analysis of the Feynman diagrams.

The tree-level $Zb\bar{b}$ interactions are contained in the terms from the Lagrangian:
\begin{eqnarray}
{\cal L}_{bb\rm NC}^{\rm tree}&=&\half\, \bar{b}_L\gamma_\mu b_L\left[-\left(g_{0L}W_{0L\mu}^{3\mu}-g_{0R}B_{0R}^{\mu}\right)\right.\nonumber\\
&&\left.\qquad\qquad +\left(g_{0L}W_{0L\mu}^{3\mu}-g_{1L}W_{1L\mu}^{3\mu}+g_{0R}B_{0R}^{\mu}-g_{1R}W_{1R}^{3\mu}\right)\kappa_A\right]\label{eq:treeZbb}\\
&&-\twothird\,\bar{b}\gamma_\mu b\left[ g_{0R}B_{0R}^{\mu}\right]\ ,\nonumber
\end{eqnarray}
where we have included the $\kappa_A$ term from Eq.~(\ref{eq:kappabb}), but we have set the $\kappa_{C_i}$'s in that equation to zero.  We shall assume $\kappa_A$ to be of similar size to the loop corrections; therefore we only include it at
leading order.

\begin{figure}[t]
\epsfig{file=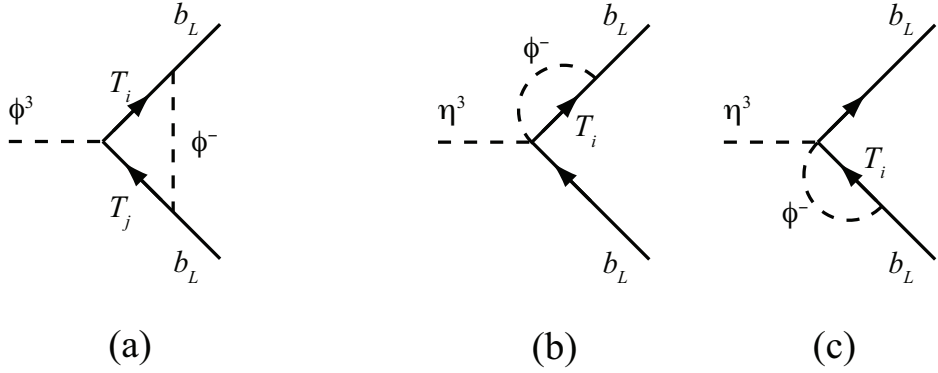,width=0.75\textwidth}
\caption{(a) Feynman diagrams for $\phi^3b\bar{b}$ correction. (b,c) Feynman diagrams for $\eta^3b\bar{b}$
correction.
\label{fig:pibb}}
\end{figure}

We have calculated the one-loop $Zb\bar{b}$ corrections using two distinct methods, both of which give the same result.  The first method is
to calculate the interaction of the Goldstone bosons with the bottom quarks in the strictly gaugeless limit,
and then use Ward identities to relate this to the $Zb\bar{b}$ corrections~\cite{Lytel:1980zh}-\cite{Abe:2009ni}.  The relevant Feynman diagrams are shown in Fig.~\ref{fig:pibb}.  Note that all analogous diagrams with other choices of Goldstone boson fields, as well as diagrams involving a three-scalar vertex, are identically zero.  We find that the diagram of Fig.~\ref{fig:pibb}(a) generates a $\phi^3b\bar{b}$ interaction,
which is finite in 4 dimensions, while the diagrams of Fig.~\ref{fig:pibb}(b,c) generate an $\eta^3b\bar{b}$ interaction,
which is divergent in dimensional regularization with $d=4-2\epsilon$.  The one-loop effective Lagrangian
that is generated by these diagrams is
\begin{eqnarray}
\Delta{\cal L}^{\rm 1-loop}_{\pi^3 b\bar{b}}&=& \bar{b}_L\gamma_\mu b_L\left[\left(\frac{\partial^\mu\phi^3}{f(2sc)}\right)\epsilon_1
+\left(\frac{\partial^\mu\eta^3}{f}\right)\epsilon_2\right]\ ,\label{eq:pibb}
\end{eqnarray}
where
\begin{eqnarray}
\epsilon_1&=&\frac{\lambda_t^2}{(4\pi)^2}-\frac{4\lambda_t^2}{(4\pi)^2}s^2\left[1+\frac{1}{2}\frac{\lambda_t^2}{\lambda_A^2}-2
\frac{\lambda_t^4}{\lambda_1^2\lambda_A^2}+\left(\frac{\lambda_t^4}{\lambda_1^2\lambda_A^2}-\frac{3}{4}\frac{\lambda_t^2}{\lambda_A^2}
\right)\ln\frac{m_{T_A}^2}{m_t^2}\right]
\ ,
\end{eqnarray}
and
\begin{equation}
\epsilon_2\ =\ \frac{\lambda_1^2}{4(4\pi)^2}
\left(\frac{1}{\epsilon}-\gamma+\frac{3}{2}-\ln{\frac{m_{T_A}^2}{4\pi\mu^2}}\right)+
\frac{\lambda_t^2}{4(4\pi)^2}
\ln{\frac{m_{T_A}^2}{m_t^2}}
\ .
\end{equation}
 We keep $\epsilon_2$ only to ${\cal O}(s^0)$, because we shall see that this term still modifies the $Zb\bar{b}$
 vertex at ${\cal O}(s^2)$.

The terms in the effective Lagrangian, Eq.~(\ref{eq:pibb}), are related to the bottom quark neutral currents, because of the
mixing of the Goldstone bosons with the neutral gauge bosons, given by
\begin{eqnarray}
{\cal L}_{\rm mixing}&=&-\frac{f}{2}(2sc)\,\partial_\mu \phi^{3}\left(g_{0L}W_{0L}^{3\mu}-g_{0R}B_{0R}^\mu\right)\\
&&
-\frac{f}{2}\,\partial_\mu \eta^{3}\left(g_{0L}W_{0L}^{3\mu}+g_{0R}B_{0R}^\mu-g_{1L}W_{1L}^{3\mu}-g_{1R}W_{1R}^{3\mu}\right)
+\cdots\ ,\nonumber
\end{eqnarray}
Following Refs.~\cite{Lytel:1980zh}-\cite{Abe:2009ni}, we can use a Ward identity to obtain the correction to the bottom quark neutral currents,
\begin{eqnarray}
\Delta{\cal L}_{bb\rm NC}^{\rm 1-loop}&=&\half \bar{b}_L\gamma_\mu b_L\left[\left(g_{0L}W_{0L\mu}^{3\mu}-g_{0R}B_{0R}^{\mu}\right)\epsilon_1\right.\label{eq:correction}\\
&&\left.\qquad\qquad +\left(g_{0L}W_{0L\mu}^{3\mu}-g_{1L}W_{1L\mu}^{3\mu}+g_{0R}B_{0R}^{\mu}-g_{1R}W_{1R}^{3\mu}\right)\epsilon_2\right]\ .\nonumber
\end{eqnarray}

\begin{figure}[t]
\epsfig{file=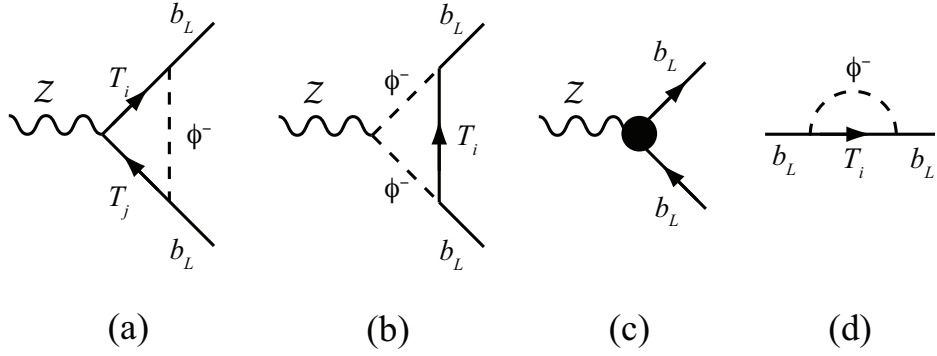,width=0.75\textwidth}
\caption{(a,b,c) Feynman diagrams for $Zb\bar{b}$ correction at one loop.  (d) Feynman diagram needed to determine
the bottom quark wave-function renormalization factor, used in the counterterm diagram (c).
\label{fig:Zbb}}
\end{figure}

We have also calculated the corrections to the $b\bar{b}$ neutral currents directly in a renormalizable $R_\xi$ gauge, but taking the limit as the gauge couplings become small.  In this limit, the leading contributions come from the diagrams of Fig.~\ref{fig:Zbb}, which are equivalent to a calculation where
the gauge fields are treated as background classical fields, while the zero-mass Goldstone bosons are dynamical quantum fields.
We obtain the exact same results in this method.

The term proportional to $\epsilon_1$ in Eq.~(\ref{eq:correction}) agrees with the SM result up to corrections of ${\cal O}(s^2)$.  The combination of gauge fields that is multiplied by $\epsilon_2$ contains the $Z$ boson at ${\cal O}(s^2)$, so we are justified
in keeping $\epsilon_2$ only to ${\cal O}(s^0)$.
To obtain the deviations from the SM prediction, we
subtract off this SM result, and use the $\overline{\rm MS}$-scheme at one-loop to define the coefficient $\hat{\kappa}_A=\kappa_A+\frac{\lambda_1^2}{4(4\pi)^2}\left(\frac{1}{\epsilon}-\gamma+\ln{4\pi}\right)$.  This gives  the final result for the $Zb\bar{b}$ vertex:
\begin{eqnarray}
\Delta{\cal L}_{bb\rm NC}^{\rm renorm}&=&\half \bar{b}_L\gamma_\mu b_L\left[\left(g_{0L}W_{0L\mu}^{3\mu}-g_{0R}B_{0R}^{\mu}\right)s^2\hat{\epsilon}_1\right.\label{eq:correction2}\\
&&\left.\qquad\qquad +\left(g_{0L}W_{0L\mu}^{3\mu}-g_{1L}W_{1L\mu}^{3\mu}+g_{0R}B_{0R}^{\mu}-g_{1R}W_{1R}^{3\mu}\right)\hat{\epsilon}_2\right]\ ,\nonumber
\end{eqnarray}
with
\begin{eqnarray}
s^2\hat{\epsilon}_1&=&\epsilon_1-\frac{2\sqrt{2}G_Fm_t^2}{(4\pi)^2}\nonumber\\
&=&\epsilon_1-\frac{\lambda_t^2}{(4\pi)^2}\left[1+4s^2\left(-\frac{\lambda_t^2}{\lambda_1^2}+\frac{1}{2}\frac{\lambda_t^4}{\lambda_A^2\lambda_1^2}\right)\right]\,,
\end{eqnarray}
so
\begin{eqnarray}
\hat{\epsilon}_1&=&-\frac{4\lambda_t^2}{(4\pi)^2}\left[\frac{3}{2}
\frac{\lambda_t^4}{\lambda_A^4}+\left(\frac{\lambda_t^4}{\lambda_1^2\lambda_A^2}-\frac{3}{4}\frac{\lambda_t^2}{\lambda_A^2}
\right)\ln\frac{m_{T_A}^2}{m_t^2}\right]
\ ,
\end{eqnarray}
and
\begin{equation}
\hat{\epsilon}_2\ =\ \hat{\kappa}_A+\frac{\lambda_1^2}{4(4\pi)^2}
\left(\frac{3}{2}-\ln{\frac{m_{T_A}^2}{\mu^2}}\right)+
\frac{\lambda_t^2}{4(4\pi)^2}
\ln{\frac{m_{T_A}^2}{m_t^2}}
\ .
\end{equation}

\begin{figure}[t]
\centerline{%
		\parbox{.30\textwidth}{\centerline{\epsfig{file=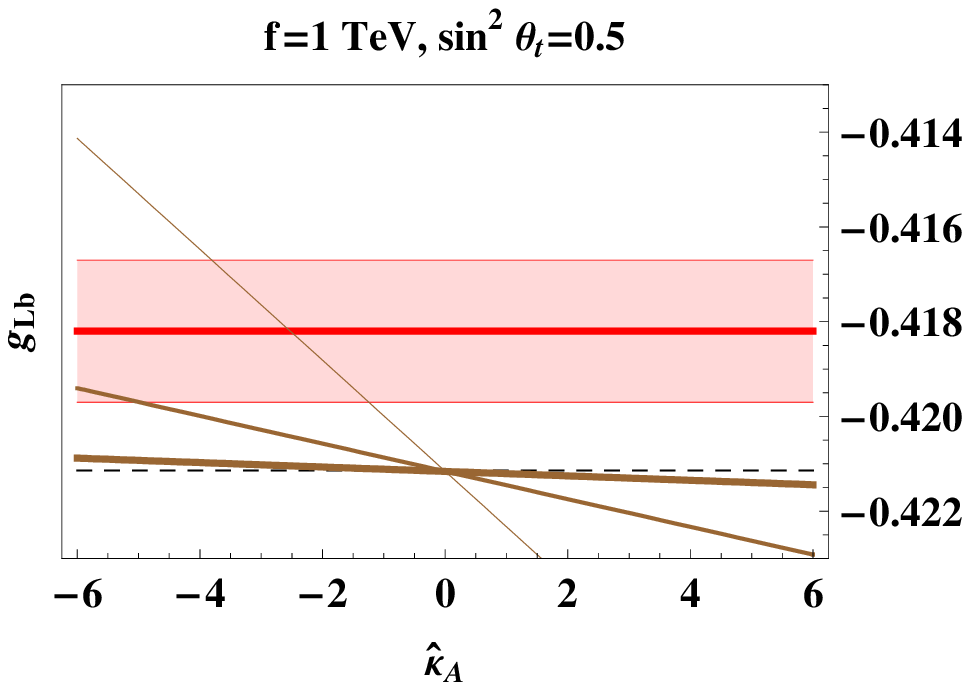,width=0.35\textwidth}}}~~~~
		\parbox{.30\textwidth}{\centerline{\epsfig{file=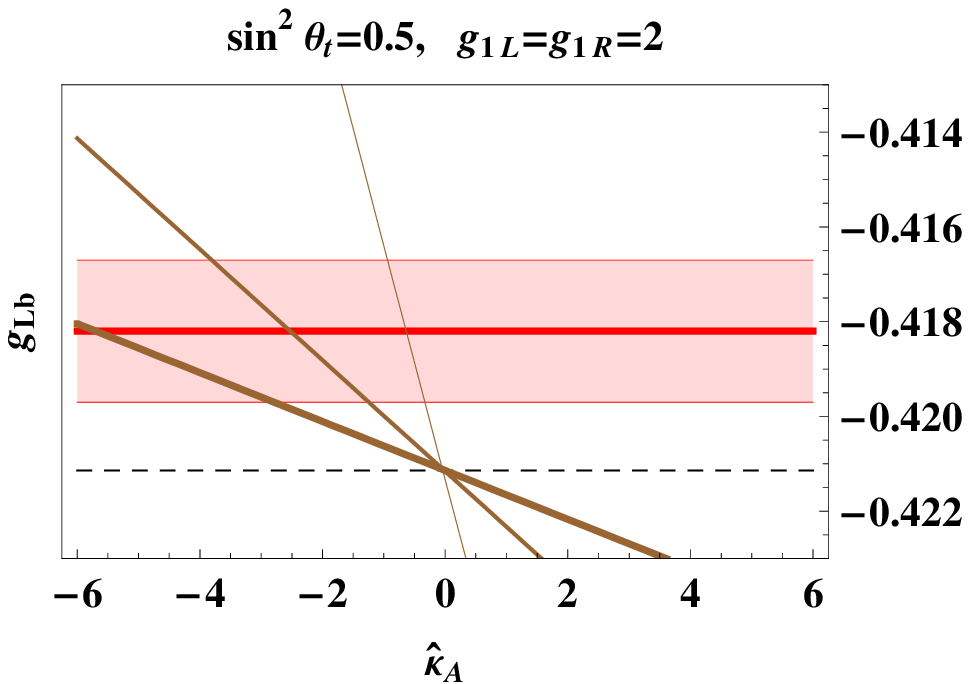,width=0.35\textwidth}}}
		\parbox{.35\textwidth}{\centerline{\epsfig{file=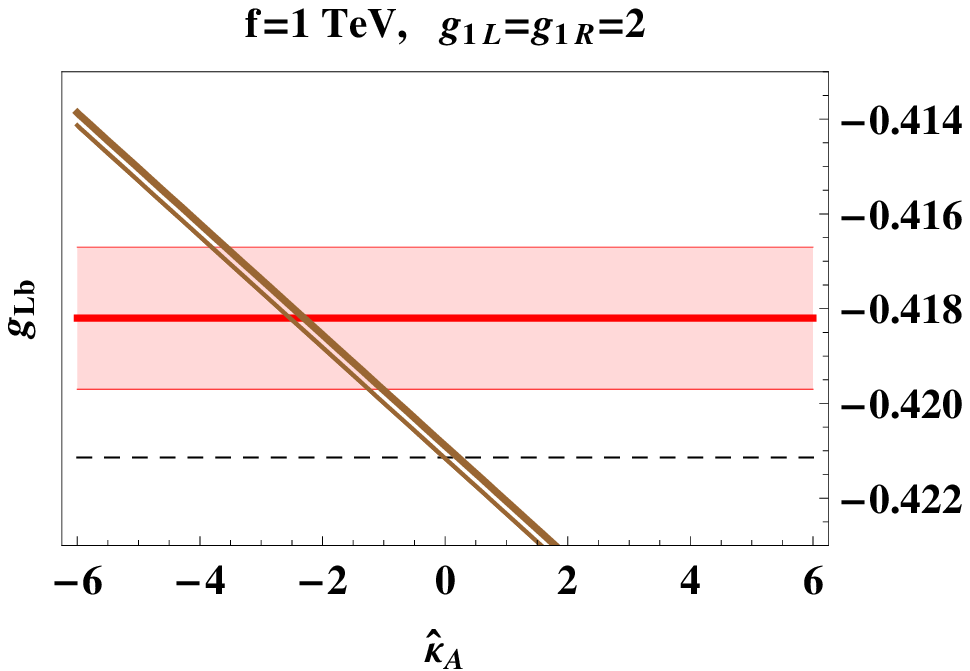,width=0.35\textwidth}}}}
\caption{Corrections to the $Zb\bar{b}$ coupling $g_{Lb}$ as a function of $\hat{\kappa}_A$ for fixed
values of $f$, $g_{1L}=g_{1R}$, and $\sin^2\theta_t$, as compared to the $\pm1\sigma$ band from experiment.
Parameter values for the different lines in each plot are given in the main text.
\label{fig:gLb}}
\end{figure}

Expanding the gauge bosons in terms of the mass eigenstates, we find the correction to the $Zb\bar{b}$ current at ${\cal O}(s^2)$ to be
\begin{eqnarray}
\Delta{\cal L}_{Zb\bar{b}}^{\rm renorm}&=&\frac{e}{\sin{2\theta}} \bar{b}_L\gamma_\mu b_LZ^\mu\ s^2\left[\hat{\epsilon}_1 +\left(\frac{g_L^2}{g_{0L}^2}-\frac{g_L^2}{g_{1L}^2}-\frac{g_R^2}{g_{0R}^2}+\frac{g_R^2}{g_{1R}^2}\right)\hat{\epsilon}_2\right]\ .
\end{eqnarray}
In Fig.~\ref{fig:gLb} we show the experimental $\pm1\sigma$ band for the left-handed $Zb\bar{b}$ coupling $g_{Lb}$, together with the SM prediction (dashed horizontal line), and our model prediction. The latter is shown as a function of $\hat{\kappa}_A$, for fixed values of $f$, $g_{1L}=g_{1R}$, and $\sin^2\theta_t$, and with the renormalization scale set to $\mu^2=m_t^2$. In Fig.~\ref{fig:gLb} (left) we set $f=1$ TeV, $\sin^2\theta_t=0.5$, and $g_{1L}=g_{1R}=2$, 4, and 10 with the line thickness increasing with $g_{1L}=g_{1R}$. The correction to $g_{Lb}$ is dominated by the renormalized tree-level term,
$\hat{\kappa}_A$, which is contained in $\hat{\epsilon}_2$. Its contribution vanishes in the limit of large $g_{1L}=g_{1R}$, as does the coefficient of $\hat{\epsilon}_2$. This leaves us with the one-loop contribution contained in $\hat{\epsilon}_1$, which, as the figure shows, is very small. In Fig.~\ref{fig:gLb} (center) we set $\sin^2\theta_t=0.5$, $g_{1L}=g_{1R}=2$, and $f=0.5$, 1, and 1.5 TeV, with the line thickness increasing with $f$. As $f$ grows, the whole nonstandard contribution to $g_{Lb}$ vanishes. Finally, in Fig.~\ref{fig:gLb} (right) we set $f=1$ TeV, $g_{1L}=g_{1R}=2$, and $\sin^2\theta_t=0.2$, 0.5, and 0.8, with the line thickness increasing with $\sin^2\theta_t$.  The dependence on $\sin^2\theta_t$ is very weak, relative to
the SM contribution.  Indeed, the lines for $\sin^2\theta_t=0.2$ and 0.5 are indistinguishable on this plot, while for $\sin^2\theta_t=0.8$ and $\hat{\kappa}_A=0$ there is a small positive one-loop correction. In general, we can conclude that the one-loop correction to $g_{Lb}$ is small, but the tree-level operator of Eq.~(\ref{eq:kappas}) can give a positive contribution to $g_{Lb}$ for negative values of $\hat{\kappa}_A=0$.

\section{Conclusions}
\label{sec:conclusions}
In Ref.~\cite{Foadi:2010bu} we introduced a LH model based on the approximate global symmetry breaking pattern $SO(5)_0\times SO(5)_1\to SO(5)$, and motivated by the deconstruction of the five-dimensional $SO(5)$ gauge-Higgs model~\cite{Medina:2007hz}. In our ``two-site'' model, the $SU(2)_{0L}\times U(1)_{0R}\times SU(2)_{1L}\times SU(2)_{1R}$ subgroup of $SO(5)_0\times SO(5)_1$ is gauged, and the spontaneous breaking to diagonal $SO(5)$ breaks the full gauge group to the electroweak symmetry group, $SU(2)_L\times U(1)_Y$. Of the ten Goldstone bosons produced, six are eaten by two iso-triplets of heavy gauge bosons, whereas the remaining four have the right quantum numbers to form an electroweak Higgs doublet. Due to vacuum misalignment of the Higgs boson potential, it acquires a vacuum expectation value, which breaks the electroweak symmetry to electromagnetism. A collective symmetry breaking mechanism ensures that no quadratic divergences are generated at one loop, leading to a natural hierarchy between the electroweak symmetry breaking scale, and the $SO(5)_0\times SO(5)_1\to SO(5)$ symmetry breaking scale.

Besides the radiatively-generated potential and the collective symmetry breaking mechanism, which are common to all LH models, our model presents a very interesting and unique feature: a stable Higgs potential can be {\em entirely} generated at one loop, both in its negative mass-squared and in the quartic coupling. In other models (for example, the littlest Higgs model~\cite{Arkani-Hamed:2002qy}), additional effective operators in the Lagrangian are necessary to generate a sufficiently large quartic Higgs coupling. In our model, instead, the presence of a larger fermion sector (which is dictated by symmetry) allows for a sufficiently small mass-squared, relative to the quartic Higgs coupling, to obtain the electroweak scale without any ad-hoc addition to the effective theory. Furthermore, the small (negative) mass-squared naturally leads to a light Higgs boson for a large range of values in the parameter space, a fact that is rather insensitive to the details of the UV completion.

In this paper we have analyzed the constraints on our model from precision electroweak measurements.
We have computed the corrections to the precision electroweak observables both at tree-level and at one loop in the fermion sector.  The tree level corrections arise due to mixing of the SM gauge bosons with heavy gauge bosons, while the dominant one-loop corrections arise from loops involving the third generation quarks, as well as their heavy partners.  As a first
approximation, we have neglected the loop contributions from the gauge sector on the assumption that they will be suppressed by roughly $m_W^2/m_t^2\simeq 0.2$, relative to the loops from the top quark sector.
We found that the tree-level corrections to $\hat{S}$ are positive and can be rather large for small values of the $SO(5)_0\times SO(5)_1$ symmetry breaking scale $f$, and/or small values of $g/g_1$, where $g$ is the weak coupling and $g_1$ is the coupling associated to the heavy gauge bosons. The one-loop corrections to $\hat{S}$ are also positive, although typically small, due to the vector-like nature of the heavy fermions.
 At tree-level there is no correction to $\hat{T}$, because of an approximate custodial symmetry. However the one-loop corrections to $\hat{T}$ are negative, a rather general feature in models with fermions which contain a bi-doublet under $SU(2)_{1L}\times SU(2)_{1R}$~\cite{SekharChivukula:2009if,Chivukula:2011jh}. The value of $\hat{T}$ depends on $f$ and the mixing between the top quark and its heavy counterparts, parametrized by $\sin\theta_t$. Specifically $-\hat{T}$ grows as $f$ decreases and/or $\sin\theta_t$ grows.

In addition to $\hat{S}$ and $\hat{T}$, we considered also the universal non-oblique corrections to the electroweak observables, $Y$ and $W$ of Barbieri {\it et al}~\cite{Barbieri:2004qk}.   At tree level, the corrections to $Y$ and $W$ scale like $g^4/g_1^4$, whereas the corrections to $\hat{S}$ scale like $g^2/g_1^2$.   Given this fact, we found that the constraints from $\hat{S}$
dominate for experimentally-allowed values of $g_1$ (except for large $f$), and therefore it is reasonable to neglect
$Y$ and $W$ for the range of parameter space in which we are interested.  We thus conclude that the universal corrections to the electroweak parameters in our theory are essentially oblique, dominated by $\hat{S}$ and $\hat{T}$. As shown in Fig.~\ref{fig:looplimits}, the experimental constraints require either $f\gtrsim 2$ TeV or else a very small value of $\sin\theta_t$. The former is more viable than the latter, since a small $\sin\theta_t$ would imply a large value of the coupling $\lambda_A$.  In addition,
as seen in Ref.~\cite{Foadi:2010bu}, the one-loop potential predicts a Higgs boson mass that tends to increase rapidly for $\sin^2\theta_t\lesssim0.2$, but increases only modestly with an  increase in the scale $f$.
 As shown in Fig.~\ref{fig:STloop}, taking $f$ to be about $2-3$ TeV predicts allowed values of $\hat{S}$ and $\hat{T}$ (at the 95\% CL) for a large range of $\sin\theta_t$. For these values of $f$, the Higgs boson can still be reasonably light with only a moderate increase in fine tuning.

In addition to the universal corrections we also considered non-universal corrections. In particular we computed the one-loop contribution to the left-handed $Zb\bar{b}$ coupling. This turns out to be cutoff-dependent; the corresponding divergence is absorbed by the dimension-four operator proportional to $\kappa_A$, in Eq.~(\ref{eq:kappas}). We found the one-loop contribution to be negligibly small, and thus the tree-level correction proportional to the renormalized $\kappa_A$ to be dominant. We also computed the correction to the $WWZ$ vertex, and found this to be negligibly small.

\begin{acknowledgments}

This work was supported by the US National
Science Foundation under grant PHY-0555544.
\end{acknowledgments}

%%%%%%%%%%%%%%%%%%%%%%%%%%%%%%%%%%%%%%%%%%%%%%%%%%%%%%

\end{document}